\documentclass[longauth]{aa}

\usepackage{graphicx}
\usepackage{lscape}
\usepackage{longtable}
\usepackage{soul}
\usepackage{lipsum}
\usepackage{hyperref}

\usepackage{chngcntr}

\usepackage{txfonts}

\newcommand{\orcid}[1]{\unskip\protect\href{https://orcid.org/#1}{\protect\includegraphics[width=8pt,clip]{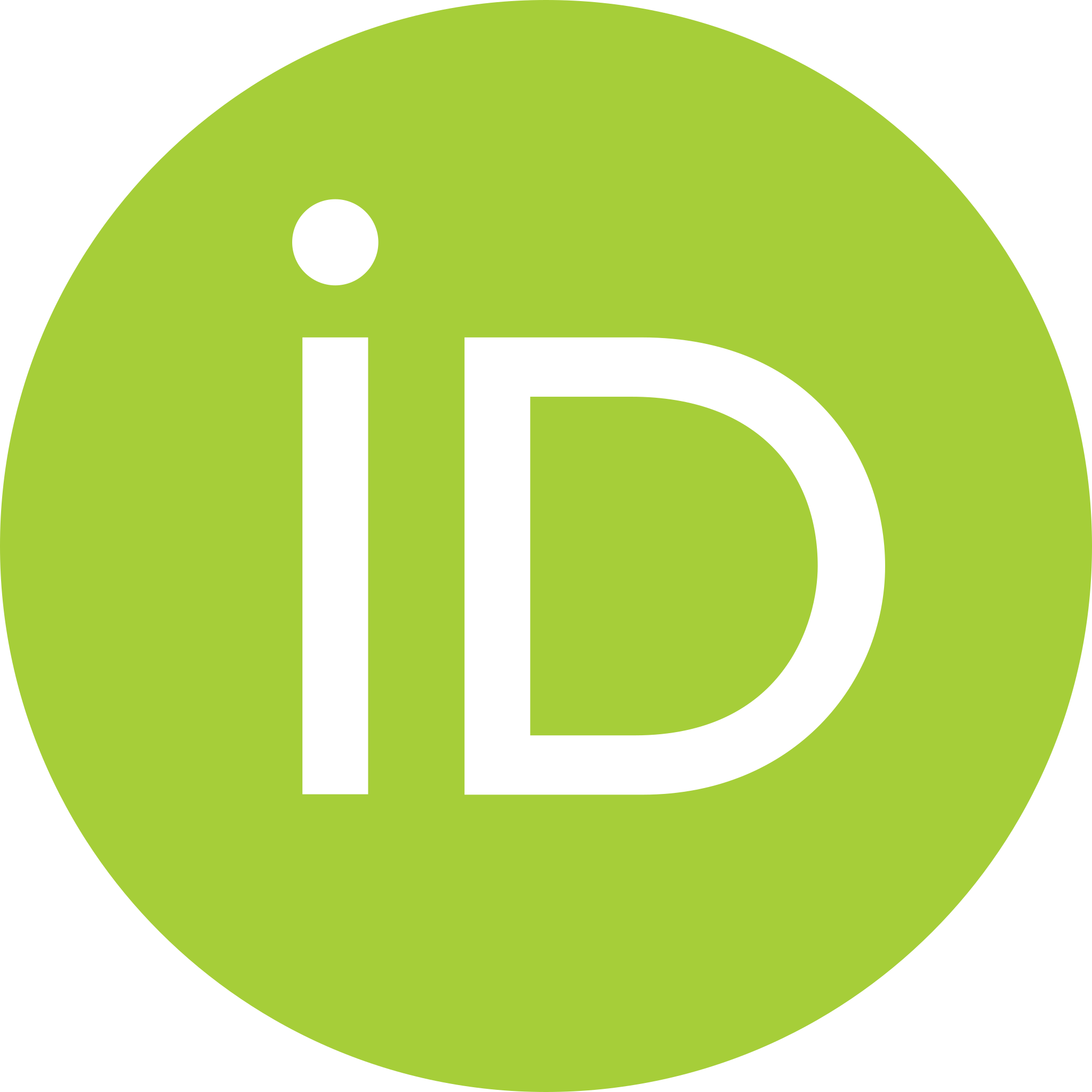}}}

\usepackage{float}
\usepackage{multirow} 

\newcommand{\Lsun}{$L_{\odot}$}
\newcommand{\Msun}{$M_{\odot}$}

\newcommand{\Rsun}{$R_{\odot}$}
\newcommand{\Mdot}{$\dot{M}$}

\newcommand{\msunyr}{$\rm{M_{\sun} \, yr^{-1}}$}

\newcommand{\mic}{$\mu$m}

\newcommand{\rstar}{$R_{*}$}
\newcommand{\mstar}{$M_{*}$}

\newcommand{\Lstar}{$L_{*}$}

\usepackage{xcolor}

\begin{document}

\title{MINDS. A multi-instrument investigation into the molecule-rich JWST-MIRI spectrum of the DF Tau binary system}
\titlerunning{MINDS. The DF Tau binary system}

\author{
Sierra L. Grant\orcid{0000-0002-4022-4899}\inst{1}   
\and Nicolas T. Kurtovic\orcid{0000-0002-2358-4796}\inst{1}  
\and Ewine F. van Dishoeck\orcid{0000-0001-7591-1907}\inst{2,1} 
\and Thomas Henning\orcid{0000-0002-1493-300X}\inst{3} 
\and Inga Kamp\orcid{0000-0001-7455-5349}\inst{4} 
\and Hugo Nowacki\inst{5}
\and Karine Perraut\inst{5}
\and Andrea Banzatti\orcid{0000-0003-4335-0900}\inst{6}
\and Milou Temmink\orcid{0000-0002-7935-7445}\inst{2}
\and Valentin Christiaens\orcid{0000-0002-0101-8814}\inst{7,8}
\and Matthias Samland\orcid{0000-0001-9992-4067}\inst{3}
\and Danny Gasman\orcid{0000-0002-1257-7742}\inst{9}
\and Beno\^{i}t Tabone\inst{10}
\and Manuel G\"udel\orcid{0000-0001-9818-0588}\inst{11,12}
\and Pierre-Olivier Lagage\inst{13}
\and Aditya M. Arabhavi\orcid{0000-0001-8407-4020}\inst{4}
\and David Barrado\orcid{0000-0002-5971-9242}\inst{14}
\and Alessio Caratti o Garatti\orcid{0000-0001-8876-6614}\inst{15,16}
\and Adrian M. Glauser\orcid{0000-0001-9250-1547}\inst{12}
\and Hyerin Jang\orcid{0000-0002-6592-690X}\inst{17}
\and Jayatee Kanwar\orcid{0000-0003-0386-2178}\inst{4,18,19}
\and Fred Lahuis\inst{20}
\and Maria Morales-Calder\'on\orcid{0000-0001-9526-9499}\inst{14}
\and G\"oran Olofsson\orcid{0000-0003-3747-7120}\inst{21}
\and Giulia Perotti\orcid{0000-0002-8545-6175}\inst{3}
\and Kamber Schwarz\orcid{0000-0002-6429-9457}\inst{3}
\and Marissa Vlasblom\orcid{0000-0002-3135-2477}\inst{2}
\and Rebeca Garcia Lopez\orcid{0000-0002-2144-0991}\inst{22}
\and Feng Long\orcid{0000-0002-7607-719X}\inst{23}\thanks{NASA Hubble Fellowship Program Sagan Fellow}
} 

\institute{
Max-Planck-Institut f\"ur Extraterrestrische Physik, Giessenbachstrasse 1, D-85748 Garching, Germany, \email{sierrag@mpe.mpg.de} 
\and 
Leiden Observatory, Leiden University, P.O. Box 9513, 2300 RA Leiden, the Netherlands 
\and
Max-Planck-Institut f\"{u}r Astronomie (MPIA), K\"{o}nigstuhl 17, 69117 Heidelberg, Germany
\and 
Kapteyn Astronomical Institute, Rijksuniversiteit Groningen, Postbus 800, 9700AV Groningen, The Netherlands
\and 
Univ. Grenoble Alpes, CNRS, IPAG, 38000 Grenoble, France
\and Department of Physics, Texas State University, 749 North Comanche Street, San Marcos, TX 78666, USA 
\and Institute of Astronomy, KU Leuven, Celestijnenlaan 200D, 3001 Leuven, Belgium
\and STAR Institute, Universit\'e de Li\`ege, All\'ee du Six Ao\^ut 19c, 4000 Li\`ege, Belgium
\and Institute of Astronomy, KU Leuven, Celestijnenlaan 200D, 3001 Leuven, Belgium
\and Universit\'e Paris-Saclay, CNRS, Institut d’Astrophysique Spatiale, 91405, Orsay, France
\and Dept. of Astrophysics, University of Vienna, T\"urkenschanzstr. 17, A-1180 Vienna, Austria
\and ETH Z\"urich, Institute for Particle Physics and Astrophysics, Wolfgang-Pauli-Str. 27, 8093 Z\"urich, Switzerland
\and Universit\'e Paris-Saclay, Universit\'e Paris Cit\'e, CEA, CNRS, AIM, F-91191 Gif-sur-Yvette, France
\and Centro de Astrobiolog\'ia (CAB), CSIC-INTA, ESAC Campus, Camino Bajo del Castillo s/n, 28692 Villanueva de la Ca\~nada, Madrid, Spain
\and INAF – Osservatorio Astronomico di Capodimonte, Salita Moiariello 16, 80131 Napoli, Italy
\and Dublin Institute for Advanced Studies, 31 Fitzwilliam Place, D02 XF86 Dublin, Ireland
\and Department of Astrophysics/IMAPP, Radboud University, PO Box 9010, 6500 GL Nijmegen, The Netherlands
\and Space Research Institute, Austrian Academy of Sciences, Schmiedlstr. 6, A-8042, Graz, Austria
\and TU Graz, Fakultät für Mathematik, Physik und Geodäsie, Petersgasse 16 8010 Graz, Austria
\and SRON Netherlands Institute for Space Research, PO Box 800, 9700 AV, Groningen, The Netherlands
\and Department of Astronomy, Stockholm University, AlbaNova University Center, 10691 Stockholm, Sweden
\and School of Physics, University College Dublin, Belfield, Dublin 4, Ireland
\and Lunar and Planetary Laboratory, University of Arizona, Tucson, AZ 85721, USA
}

\abstract
{The majority of young stars form in multiple systems, the properties of which can significantly impact the evolution of any circumstellar disks.} 
{We investigate the physical and chemical properties of the equal-mass, small-separation ($\sim$66 milliarcsecond, $\sim$9 au) binary system DF Tau. Previous spatially resolved observations indicate that only DF Tau A has a circumstellar disk, while DF Tau B does not, as concluded by a lack of accretion signatures and a near-infrared excess.}
{We present JWST-MIRI MRS observations of DF Tau. The MIRI spectrum shows emission from a forest of H$_2$O lines and emission from CO, C$_2$H$_2$, HCN, CO$_2$, and OH. Local
thermodynamic equilibrium slab models were used to determine the properties of the gas. The binary system is not spatially or spectrally resolved in the MIRI observations; therefore, we analyzed high spatial and spectral resolution observations from ALMA, VLTI-GRAVITY, and IRTF-iSHELL to aid in the interpretation of the molecular emission observed with JWST.}
{The 1.3 mm ALMA observations show two equal-brightness sources of compact ($R\lesssim$3 au) continuum emission that are detected at high significance, with separations consistent with astrometry from VLTI-GRAVITY and movement consistent with the known orbital parameters of the system. We interpret this as a robust detection of the disk around DF Tau B, which we suggest may host a small ($\sim$1 au) cavity; such a cavity would reconcile all of the observations of this source. In contrast, the disk around DF Tau A is expected to be a full disk, and spatially and spectrally resolved dust and gas emission traced by ground-based infrared observations point to hot, close-in ($\lesssim0.2$ au) material around this star. 
High-temperature emission ($\sim$500-1000 K)  from H$_2$O, HCN, and potentially C$_2$H$_2$ in the MIRI data likely originates in the disk around DF Tau A, while a cold H$_2$O component  ($\lesssim$200 K) with an extended emitting area is consistent with an origin from both disks.}
{Given the unique characteristics of this binary pair, complementary observations are critical for constraining the properties of these disks. Despite the very compact outer disk properties, the inner disk composition and the conditions of the DF Tau disks are remarkably similar to those of isolated systems, suggesting that neither the outer disk evolution nor the close binary nature are driving factors in setting the inner disk chemistry in this system. However, constraining the geometry of the disk around DF Tau B, via higher angular resolution ALMA observations for instance, would provide additional insight into the properties of the mid-infrared gas emission observed with MIRI. JWST observations of spatially resolved binaries, at a range of separations, will be important for understanding the impact of binarity on inner disk chemistry more generally. }


\date{Received May 17, 2024; accepted June 15, 2024}

\maketitle

\section{Introduction}\label{sec: intro}
The properties of molecular emission arising from the warm inner regions of protoplanetary disks can be signposts of disk evolution. Processes such as inner disk clearing, the radial drift of dust grains, and accretion outbursts can all impact the chemical signatures of the gas in the inner 10 au that can best be probed by mid-infrared spectroscopy (e.g., \citealt{banzatti20,ciesla_cuzzi06,molyarova18}). While mid-infrared spectroscopy is an important tool for constraining the properties of this area, complementary observations at other wavelengths can be crucial in breaking degeneracies to determine which process is dominating the observed chemical trends.

Stellar multiplicity is a common occurrence in star formation, and this multiplicity can drastically impact disk evolution in such systems. The stellar mass ratio, the binary separation, and the disk-orbit alignment are a few of the main factors that determine the evolution of circumstellar or circumbinary disks in these systems. Disk truncation in binary systems -- where the disks in an equal-mass system can have a maximum size of $\sim$1/3 of the binary separation -- has long been predicted \citep{papaloizou_pringle77}. Recent modeling has shown that dust disks in binary systems are often even smaller than the truncation radius due to efficient dust radial drift induced by the companion \citep{zagaria21a}, an effect that is also observed in ALMA observations \citep{manara19a,zagaria21b,tofflemire24}. However, despite these small, rapidly evolving disks, exoplanets are observed in binary systems (e.g., \citealt{eggenberger,marzari19,su21}).

Variability at optical wavelengths has been observed in DF Tau since the early 1900s \citep{lamzin01,zaitseva76}. The binary nature of this system was first discovered in 1989 thanks to lunar occultation measurements  \citep{chen90}. Ground-based and space-based observations in the 1990s and early 2000s allowed for the first spatially resolved observations of the emission coming from DF Tau A and B \citep{thiebaut95, ghez97, hartigan04}. Most of these optical and UV observations showed that only DF Tau A had indications of ongoing accretion, although some showed evidence for accretion around DF Tau B \citep{hartigan03}. Spatially resolved Keck observations in the early to mid 2000s showed that, in addition to lacking accretion signatures, DF Tau B also lacked (1) a near-infrared excess, out to $\sim$4 $\mu$m \citep{schaefer14,allen17}, a signature of inner disk clearing (e.g., \citealt{espaillat14}) and (2) a veiling of photospheric lines \citep{allen17,prato23}, another indicator of a lack of strong accretion (e.g., \citealt{hartigan03}). High-resolution spectra of the [OI] line in DF Tau show a complex line profile, indicative of winds and jets \citep{simon16, banzatti19}, adding to the dynamic view of this system. 

Since the discovery of DF Tau's binarity, studies have consistently found that this system is a close-separation, equal-mass binary pair. Both DF Tau A and B have spectral types of $\sim$M2 and stellar masses of 0.55 \Msun\ \citep{allen17}. With an orbital period of only 46 years \citep{allen17}, the binary separation is rapidly evolving; the separation was $\sim$66 milliarcseconds ($\sim$9 au) at the time of the observations presented in this work.

To characterize this unique system, we present JWST-MIRI, VLTI-GRAVITY, and ALMA observations with additional analysis of data from IRTF-iSHELL. By utilizing this wealth of complementary data, we gain new insight into this binary pair and shed light on the physical and chemical evolution of close-separation binaries. We present the observations in Section~\ref{sec: observations}, our analysis of the VLTI-GRAVITY and ALMA observations in Section~\ref{sec: VLTI and ALMA results}, and our analysis of the JWST-MIRI observations in Section~\ref{sec: JWST analysis}. We discuss our findings in Section~\ref{sec: discussion} and offer a summary in Section~\ref{sec: summary}.

\begin{figure*}[h]
    \centering
    \includegraphics[scale=0.57]{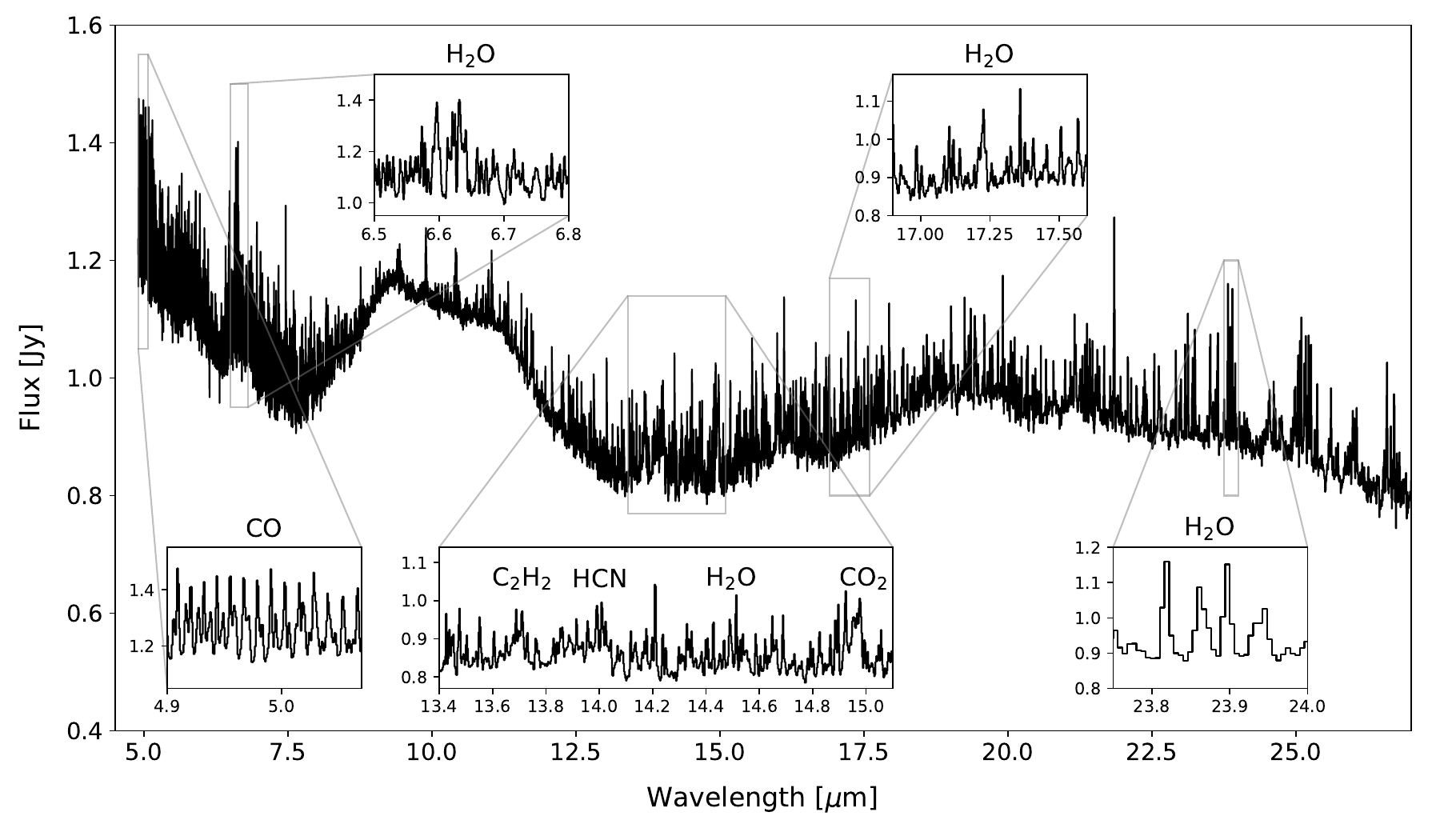}
    \caption{ JWST-MIRI MRS spectrum for DF Tau. Note the line forest present in the spectrum, which has a very high line-to-continuum ratio.  }
    \label{fig: spec}
\end{figure*}

\section{Observations and data reduction}\label{sec: observations}

\subsection{JWST-MIRI observations and data reduction}\label{subsec: obs}
DF Tau was observed with the Mid-InfraRed Instrument (MIRI; \citealt{rieke15, wright15,wright23}) in the Medium Resolution Spectroscopy (MRS; \citealt{wells15,argyriou23}) mode on 20 February 2023. These observations are part of the MIRI mid-INfrared Disk Survey (MINDS) JWST guaranteed time observation program (PID: 1282, PI: T. Henning, \citealt{kamp23,henning24}). A four-point dither was performed in the positive direction. The total exposure time was 27.2 minutes. Target acquisition was not utilized in these observations. 
DF Tau A and B are not resolved separately. Therefore, the MIRI MRS spectrum is taken to be the sum of any star and disk emission from both sources. 

The MIRI data are reduced using a hybrid pipeline\footnote{The pipeline and associated documentation are available at \url{https://github.com/VChristiaens/MINDS}} \citep{minds_notebook}, combining routines from the standard JWST pipeline \citep{bushouse_1.13.4} using CRDS context 1202, and from the VIP package \citep{GomezGonzalez2017,Christiaens2023}. The pipeline is structured around three main stages that are the same as in the JWST pipeline, namely Detector1, Spec2 and Spec3. After the first stage (Detector1), stray light is corrected using the corresponding standard JWST pipeline step and a background estimate is subtracted by using a direct pair-wise dither subtraction. For each band, the centroid of the source was identified through a 2D Gaussian fit in a positive thresholded mean spectral image. The spectrum was extracted through aperture photometry with apertures set at each centroid location, and their size set to 1.5 full width at half maximum (FWHM) to minimize self-subtraction. Fringe correction was done both on the 2D detector images and on the extracted 1D spectrum.

The JWST-MIRI spectrum is presented in Figure~\ref{fig: spec}, which shows a wealth of emission lines from CO, C$_2$H$_2$, HCN, CO$_2$, and OH and a forest of strong H$_2$O lines\footnote{The reduced spectrum that we present here is available by request from the corresponding author and will become available at \url{https://spexodisks.com}}. It is one of the most line-rich spectra in the MINDS sample.

\subsection{ALMA}

DF Tau was observed with ALMA as part of the projects 2019.1.01739.S (PI: Tofflemire) and 2021.1.00854.S (PI: Long). Each project observed DF Tau twice, once with a long baseline configuration and once with a compact baseline configuration, which we name LB and SB, respectively, as listed in Table~\ref{tab:alma_obs_log}. The observations SB1 and LB1 had the correlator configured to observe 4 spectral windows, with one of them centered at 230.538\,GHz to observe the $^{12}$CO $J$=2-1 line with a frequency resolution of 488.28\,kHz ($\sim$0.6 km s$^{-1}$), and the remaining three were centered at 232.6, 245.0 and 246.9\,GHz to observe dust continuum emission. The observations SB2 and LB2 had the correlator configured to observe with 6 spectral windows: two were centered at 219.56 (C$^{18}$O $J$=2-1) and 220.399\,GHz ($^{13}$CO $J$=2-1) with a frequency resolution of 282.23kHz; two at 230.538 ($^{12}$CO $J$=2-1) and 231.322\,GHz (N$_2$D$^+$ $J$=3-2) with a frequency resolution of 141.11kHz; and two at 218 and 233\,GHz to observe dust continuum emission. An independent analysis of the dust continuum and analysis of the $^{12}$CO data will be presented in Kutra et al. in preparation. 

The observations of both programs were pipeline-calibrated with the \texttt{scriptforPI} delivered by ALMA. Then, \texttt{CASA 5.6.2} was used to extract the dust continuum emission from all the spectral windows. As in \citet{andrews18}, we flagged the channels located at $\pm20$\,km\,s$^{-1}$ from the expected lines, using as line center the approximate system velocity at the local standard of rest (typically 5\,km\,s$^{-1}$ for the disks in the Taurus star-forming region). The remaining channels were combined into a ``pseudo-continuum'' measurement set, which we averaged into 125\,MHz channels and $6\,$s bins in order to reduce the data volume. During the LB1 observation, the antenna DV23 returned a temperature of 0\,K while observing the flux calibrator J0510+1800, and thus we removed this antenna from the self-calibration and analysis process. 

The ALMA observations were obtained at a time near periastron, when the relative motion of the binaries is at a maximum ($\sim$14 to 17 mas yr$^{-1}$, considering the orbital parameters from \citealt{allen17}). Given that both stars could potentially have a disk contributing to the millimeter emission, we self-calibrated each observation individually, as combining them would assume that the brightness distribution has not changed in the time spanned between epochs. We self-calibrated the data using the continuum emission, which was imaged with the task \texttt{tclean}, using a Briggs robust parameter of 0.5 for all epochs. To avoid introducing point spread function artifacts, we lowered the \texttt{gain} parameter to $0.02$ and increased the \texttt{cyclefactor} to $1.5$, for more conservative imaging compared to the default values\footnote{Check \url{https://casa.nrao.edu/docs/taskref/tclean-task.html} for a detailed description of the \texttt{tclean} parameters}. Each image was cleaned down to a 1$\sigma$ threshold. 
In each phase and amplitude calibration we combined all the scans and spectral windows to estimate the \texttt{gaincal} solution table. Phase calibrations were applied until no improvement was seen in the peak signal-to-noise ratio, and then only one amplitude calibration was applied to each observation. 
For further analysis, we extracted the visibility table of each observation independently. The central frequency of each channel was used to calculate its visibility coordinates in wavelength units. The calibrated ALMA data are publicly available\footnote{\url{https://zenodo.org/records/11215003}}.

\begin{figure*}
  \centering
  \includegraphics[scale=0.47]{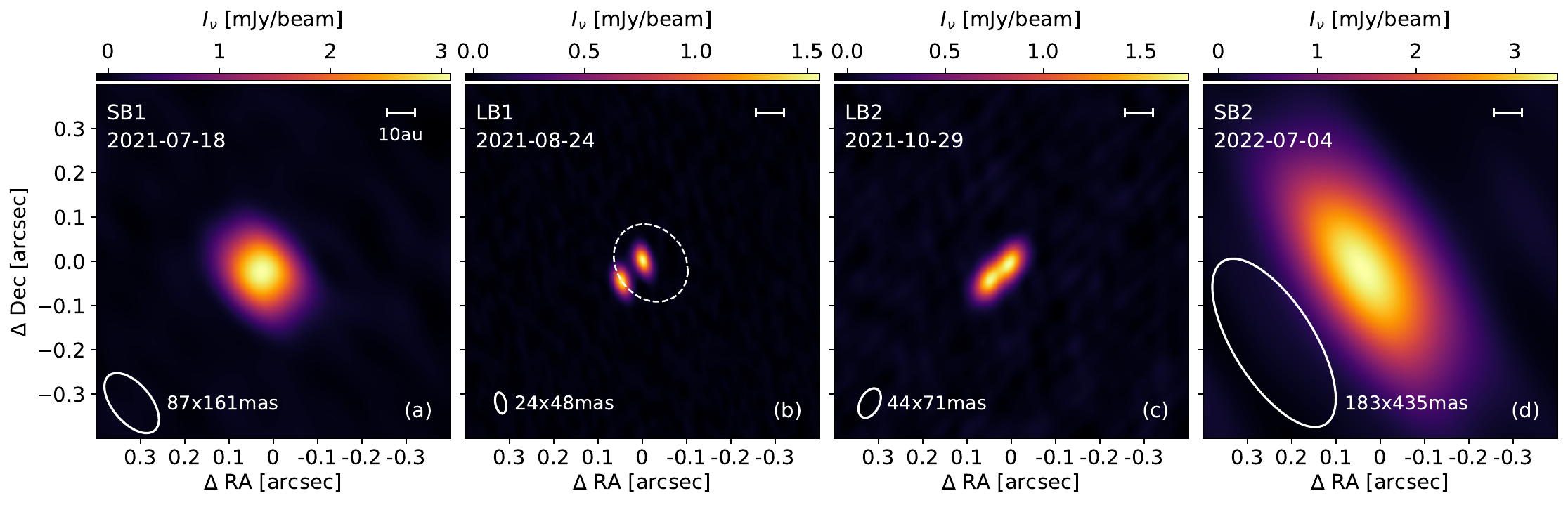}
  \caption{ ALMA dust continuum emission at each epoch, sorted by observation date. SB and LB correspond to the short-baseline and long-baseline observations, respectively. The angular resolution is represented by the white ellipse at the bottom-left corner of each panel. The scale bar is 10\,au at the distance of the source. The orbit of the system from \citet{allen17} is shown as a dotted line in panel (b) for reference. DF Tau A is at the center of the image, while DF Tau B is to the lower left.}
  \label{fig:ALMA_continuum}
\end{figure*}

\subsection{VLTI-GRAVITY}\label{subsec: gravity}
DF Tau was observed with VLTI-GRAVITY \citep{gravity17} on 28 October 2022. We recombined the four auxiliary telescopes in the medium configuration D0-G2-J3-K0, leading to a maximal angular resolution of $\lambda$/2B (with $B$ the maximal separation between the telescopes) of about 2 milliarcseconds. The target photons in the K-band (2.0 to 2.45 \mic) were split between the scientific instrument equipped with a spectrometer with a spectral resolution of 4000, and the fringe tracker instrument, which tracks the fringes at a frequency rate of $\sim$300 Hz. We recorded 6 files of 6 minutes each on the target. An interferometric calibrator, HR~1061, was also observed in the same conditions to be able to calibrate the instrument transfer function. All of the data were processed with the standard GRAVITY pipeline \citep{gravity_DRS}. The visibilities squared and the closure phases exhibit clear signatures of binarity with sinusoidal variations of visibilities as a function of spatial frequency (Figure~\ref{fig: gravity model}). The slight decrease in the visibility sinusoidal modulations at higher spatial frequencies is also indicative of a marginally resolved component such as a circumstellar and/or a circumbinary disk. The hydrogen recombination line Br$\gamma$ at 2.166 \mic\ and two helium lines at 2.01 and 2.06 \mic\ were detected in the GRAVITY spectrum. However, the data quality is not good enough to derive interferometric quantities across these lines, and therefore we focused our analysis on the K-band continuum observations.

\subsection{IRTF-iSHELL}
DF Tau was observed with iSHELL \citep{rayner16, rayner22} on the Infrared Telescope Facility (IRTF) on 24 January 2022. These data are published in \cite{banzatti23a} (see also \citealt{banzatti22a})\footnote{The spectrum is available at \url{https://spexodisks.com/ExploreData/ishell_4516nm_5232nm_vstar_df_tau}}. In these high spectral resolution M-band data (4.5 to 5.2 \mic), \cite{banzatti23a} find that the CO fundamental emission lines have two velocity components, a broad component (FWHM=113 km s$^{-1}$) and a narrow component (FWHM=49 km s$^{-1}$). Additionally, H$_2$O is detected at 5 \mic\ with a velocity similar to the broad CO component (FWHM=100 km s$^{-1}$). We note that even at the high spectral resolution of the iSHELL data ($R\sim92,000$, 3.3 km s$^{-1}$), DF Tau A and B are not spectrally resolved, given a radial velocity offset of $\sim$1 km s$^{-1}$ at the time of the observations.

\section{Constraining the physical properties of the two disks in the DF Tau system}\label{sec: VLTI and ALMA results}

\subsection{The discovery of a disk around DF Tau B with ALMA}
When imaged with ALMA at high angular resolution, the DF Tau system resolves into two blobs of dust continuum emission robustly detected at over 100$\sigma$ at peak emission. These two blobs are spatially separated in the LB1 and LB2 images, as shown in panels (b) and (c) in Figure \ref{fig:ALMA_continuum}, and the separation is detected in the visibilities of the four epochs. When paired with the orbital information from the literature (e.g., \citealt{allen17}) and the separation between DF Tau A and B from the GRAVITY data (65.9 $\pm$ 0.6 mas), it appears that the two blobs correspond to two circumstellar disks, one around DF Tau A, as expected, and one around DF Tau B, which was unexpected given the lack of accretion signatures and near-infrared excess as reported by \cite{allen17}.


We modeled the ALMA emission of the disks as described in Appendix~\ref{sec:alma_analysis}, with the main results summarized in Table \ref{tab:uv_mcmc}. At the current angular resolution, we cannot set strong constraints on the morphology of these compact disks from the 1.3\,mm dust continuum observation.  However, when modeled with axisymmetric Gaussian profiles, we find the disks to be compact ($<3\,$au in radius), and of similar size and flux. The Gaussian models for DF Tau B show a preference for non-centrally peaked brightness distributions to the $1\,\sigma$ of confidence, differently from DF Tau A, which is centrally peaked to the $1\,\sigma$ level. This potentially indicates a cavity in the disk around DF Tau B, which would be consistent with the lack of near-infrared excess around this star; however, based on the current ALMA observations, we can neither conclusively confirm nor reject the presence of a cavity in the disk around DF Tau B. However, any cavity would be $\lesssim$1.5 au in radius, as larger cavities would have been detected. We discuss this scenario in Section~\ref{subsubsec: cavity in B}.

\begin{table}[t]
\caption{Disk properties from the ALMA visibility modeling.}
\centering
\begin{tabular}{ c|c|c } 
  \hline
  \hline
\noalign{\smallskip}
           & Property      & Best value $\pm 3\sigma$  \\
\noalign{\smallskip}
  \hline
  \hline
\noalign{\smallskip}
1.3 mm Flux [mJy] & $F_{A}$   & $2.00 \pm 0.06$ \\
                  & $F_{B}$   & $2.00 \pm 0.15$ \\
\noalign{\smallskip}
  \hline
\noalign{\smallskip}
Dust Mass [$M_{\oplus}$]     & $M_{dust,A}$   & $1.18 \pm 0.16$  \\
     & $M_{dust,B}$   & $1.18 \pm 0.18$  \\
\noalign{\smallskip}
  \hline
\noalign{\smallskip}
Relative         & $\Delta\text{RA}_{\text{SB1}}$  & $-47.5 \pm 2.5 $  \\
Astrometry       & $\Delta\text{Dec}_{\text{SB1}}$ & $-46.2 \pm 2.8 $  \\
DF Tau A-B [mas] & $\Delta\text{RA}_{\text{SB2}}$  & $-69.2 \pm 21.1 $  \\
                 & $\Delta\text{Dec}_{\text{SB2}}$ & $-29.3 \pm 31.9$  \\
                 & $\Delta\text{RA}_{\text{LB1}}$  & $-47.4 \pm 0.3 $ \\
                 & $\Delta\text{Dec}_{\text{LB1}}$ & $-47.1 \pm 0.4 $  \\
                 & $\Delta\text{RA}_{\text{LB2}}$  & $-49.9 \pm 1.2 $  \\
                 & $\Delta\text{Dec}_{\text{LB2}}$ & $-44.0 \pm 1.5 $  \\
\noalign{\smallskip}
  \hline
\noalign{\smallskip}
\textit{Size}$^{*}$ [mas] & $R_{A,68\%}$ & $\leq13.8 \pm 0.5$ \\
                 & $R_{B,68\%}$ & $\leq13.9 \pm 0.7$ \\
                 & $R_{A,90\%}$ & $\leq20.1 \pm 1.0$ \\
                 & $R_{B,90\%}$ & $\leq19.6 \pm 1.2$ \\
\noalign{\smallskip}
  \hline
  \hline
\end{tabular}
\tablefoot{The disks geometries and relative astrometry are free parameters during the MCMC fitting. The disk sizes and fluxes are not directly fitted, but calculated from the best results shown in Table \ref{appendix:tab:uv_mcmc}, where the remaining free parameters are shown. Uncertainties are the 3$\sigma$ values of the distributions. Dust masses are computed assuming a dust temperature of 20 K, a wavelength of 1.33 mm, an opacity of 2.254 cm$^{2}$ g$^{-1}$ \citep{beckwith90}, and a distance of 140$\pm$10 pc (see Section~\ref{sec: JWST analysis}). $^{*}$ indicates parameters that should be taken with caution. As the emission is only marginally resolved, the values are not well constrained and the uncertainties instead reflect the uncertainties of the model and not the true uncertainties of the data.}
\label{tab:uv_mcmc}
\end{table}

\subsection{The inner dust disk around DF Tau A with VLTI-GRAVITY}

The VLTI-GRAVITY K-band (2.2 $\mu$m) interferometry allows us to constrain the separation of the two stars, DF Tau A and B, at the time of the observations and detect and constrain the inner dust disk size. Several models were tested to reproduce the VLTI-GRAVITY observations, all with two point-like sources, but with different disk contributions: (1) no disk material, (2) a Gaussian disk around one source, (3) a Gaussian disk around each source, and (4) a circumbinary disk. The circumbinary model never converged making this possibility very unlikely and the two disk model did not reproduce the data as well as the binary-only and the binary-plus-one-disk models; therefore, we only discuss models (1) and (2) here.

Following the van Cittert-Zernike theorem, we fit our dataset with a binary model composed of a point-like primary target (DF Tau A) and a point-like secondary (DF Tau B). The binary-plus-disk model is the same, but now adding a Gaussian disk model to the DF Tau A point-like source. DF Tau A is chosen as the disk-bearing source due to the lack of near-infrared excess around DF Tau B (Figure~\ref{fig: sed}). The spatial frequencies probed by our uv-plane coverage are not sufficient to constrain either the disk inclination or the flux distribution inside the disk, so it is considered flat (only a radius, given by the HWHM, is derived). Due to the degeneracy between size and fluxes for marginally resolved sources, we constrained the relative flux of each component ($F_A$, $F_B$, and $F_{disk}$) using the K-band magnitude from \cite{schaefer14}. The relative contribution of each stellar component of the binary varies from one model to the other, but scales in the same manner: the primary represents $\sim60 \%$ of the total flux, the secondary is $\sim30\%$ (consistent with the flux ratio of $\sim$2 from \citealt{schaefer14}), and the contribution of the disk is then $\sim10\%$. More details are provided in Appendix~\ref{sec: gravity model}. 

For each model, after exploring the parameter space manually to identify the region of global $\chi^2$ minimum, a Markov chain Monte Carlo (MCMC) procedure was adopted. The posterior distribution was used, assuming a $\chi^2$ statistic, to derive the final best-fit value and its associated uncertainties. The binary-plus-disk model provides a significant improvement in the $\chi^2_r$ compared to the binary-only model (from 8 to 5). This best-fit model has a half width at half maximum (HWHM) of $\sim1$ mas, corresponding to an emitting radius of $\sim$0.14 au. The best-fit parameters for the two models are given in Table~\ref{tab:GRAV_fit} and the data compared to the best-fit model are provided in Figure~\ref{fig: gravity model}. While higher angular resolution observations are needed to conclude the exact geometry and flux distribution of the circum-primary disk, the current data are well reproduced by this model.

\begin{table}
    \centering
      \caption{Best-parameter fit of the VLTI-GRAVITY data for each model considered.}
    \begin{tabular}{c|c|c}
    \hline
    \hline
        \multirow{2}{*}{Parameter} & \multicolumn{2}{c}{VLTI-GRAVITY Models} \\\cline{2-3}
         & Binary only & Binary + disk around A\\
    \hline
        $\Delta\text{RA}$ [mas] & $57.39^{+0.18}_{-0.17} $ & $57.38 \pm 0.22$ \\
        $\Delta\text{Dec}$ [mas] & $-32.35 \pm 0.34$ &  $-32.38^{+0.41}_{-0.42}$ \\
        $\rho$ [mas] & $65.9 \pm 0.5$  & $65.9 \pm 0.6$ \\
        $F_A$ & $0.66 \pm0.03$ &   $0.61^{+0.04}_{-0.06}$  \\
        $F_B$ &  $0.34 \pm 0.03$  & $0.28 \pm 0.04$   \\
        $F_{disk}$ &  -  &    $0.11 \pm 0.07$  \\
        HWHM [mas] & - &   $1.0 \pm 0.6$ \\
        HWHM [au] & - &  $0.14 \pm 0.09$  \\
        $\chi_r^2$ & 8.38  & 5.05 \\
    \hline
    \end{tabular}
    \tablefoot{Uncertainties are derived assuming a $\chi^2$ statistic for all parameters. $\Delta\text{RA}$ and $\Delta\text{Dec}$ are the offsets of DF Tau B relative to DF Tau A and $\rho$ is the separation. $F_A$ and $F_B$ correspond to the fraction of the total flux coming from DF Tau A and B, respectively. The HWHM corresponds to the disk radius.}
    \label{tab:GRAV_fit}
\end{table}

\subsection{The inner gas disk around DF Tau A with IRTF-iSHELL}\label{subsec: ishell gas}

\begin{figure}
    \centering
    \includegraphics[scale=0.54]{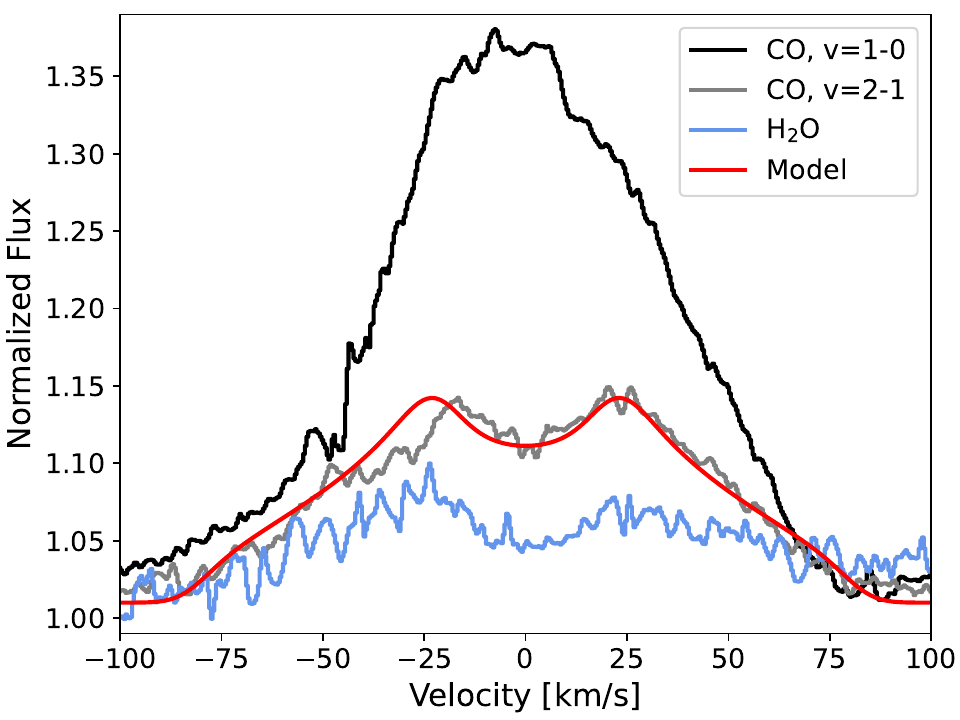}
    \caption{Stacked line profiles for the CO v=1-0 (black), CO v=2-1 (gray), and the 5 \mic\ H$_2$O (blue) lines from the IRTF-iSHELL spectrum of DF Tau \citep{banzatti23a}. A Keplerian line profile, calculated using an inclination of 67$^\circ$ and a stellar mass of 0.55 \Msun, shown in red, matches the v=2-1 lines.  }
    \label{fig: line profiles}
\end{figure}

While the VLTI-GRAVITY data trace the inner disk dust structure, we can use high spectral resolution observations in the M-band to study the inner disk gas structure via CO and H$_2$O rovibrational emission (e.g., \citealt{bast11,brown13,banzatti15b, banzatti22a}). The CO and H$_2$O line profiles for DF Tau are shown in Figure~\ref{fig: line profiles}, where it can be seen that two velocity components are present in the CO v=1-0 lines, a broad component coming from Keplerian rotation of the gas and a narrow component that may at least in part originate in a sub-Keplerian disk wind \citep{pontoppidan11b,banzatti22a}. The emitting radii of the gas can be determined by using the line widths, stellar mass, and disk inclination and assuming Keplerian rotation (see, e.g., \citealt{banzatti22a, grant24a}). The inclination can be estimated from the line profiles following \cite{banzatti15b} (see their Equation 1), which gives an inner disk inclination of $\sim$67$^\circ$. This inclination is very different from the inclination of 18.9$^\circ$ determined from ALMA; however, given that the disks are only marginally resolved with ALMA and due to the potential for disk warping in such a dynamic system, we opted to take the inclination from the CO line profiles. When this inclination is adopted, the emitting radii for the CO broad component and the H$_2$O are determined to be 0.13 and 0.17 au, respectively. This is consistent with the radius determined with VLTI-GRAVITY, as is seen in other objects as well \citep{banzatti23a}. 

In Figure~\ref{fig: line profiles}, we compare the observed CO v=2-1 line, whose double-peaked profile demonstrates Keplerian broadening by disk rotation, to a Keplerian model that assumes the line brightness to decrease with radius as $R^{- \alpha}$ between and inner and outer radius ($R_{in}$ and $R_{out}$) as done in previous work (e.g., \citealt{salyk11b,banzatti22a}). To obtain a good match, the model has $R_{in}$ = 0.06 au (which determines the line wings) and $R_{out}$ = 0.9 au (which determines the line peaks), with an exponent $\alpha = 2.3$. In this model, 90\% of the line flux is coming from 0.06 to 0.15 au. While the three parameters are slightly degenerate, providing confidence regions of $\approx 10\%$ around these values, this simple model strongly supports the conclusion that the M-band CO and H$_2$O emission comes from very small radii ($<$0.2 au). We did not attempt to match the narrow component with a Keplerian model, as it may originate from a sub-Keplerian disk wind. 

Given that the VLTI-GRAVITY data are well reproduced by a single inner disk model with a radius of 0.14 au and the high spectral resolution data show broad Keplerian emission coming from gas very close to the star leads us to conclude that DF Tau A has a very hot inner disk of gas and dust. This inner disk structure is then used to aid in the interpretation of the spectrally and spatially unresolved JWST-MIRI MRS data.

\subsection{VLTI-GRAVITY and ALMA astrometry of DF Tau B}\label{subsec:DFTauB_astrometry}

\begin{figure}
    \centering
    \includegraphics[scale=0.44]{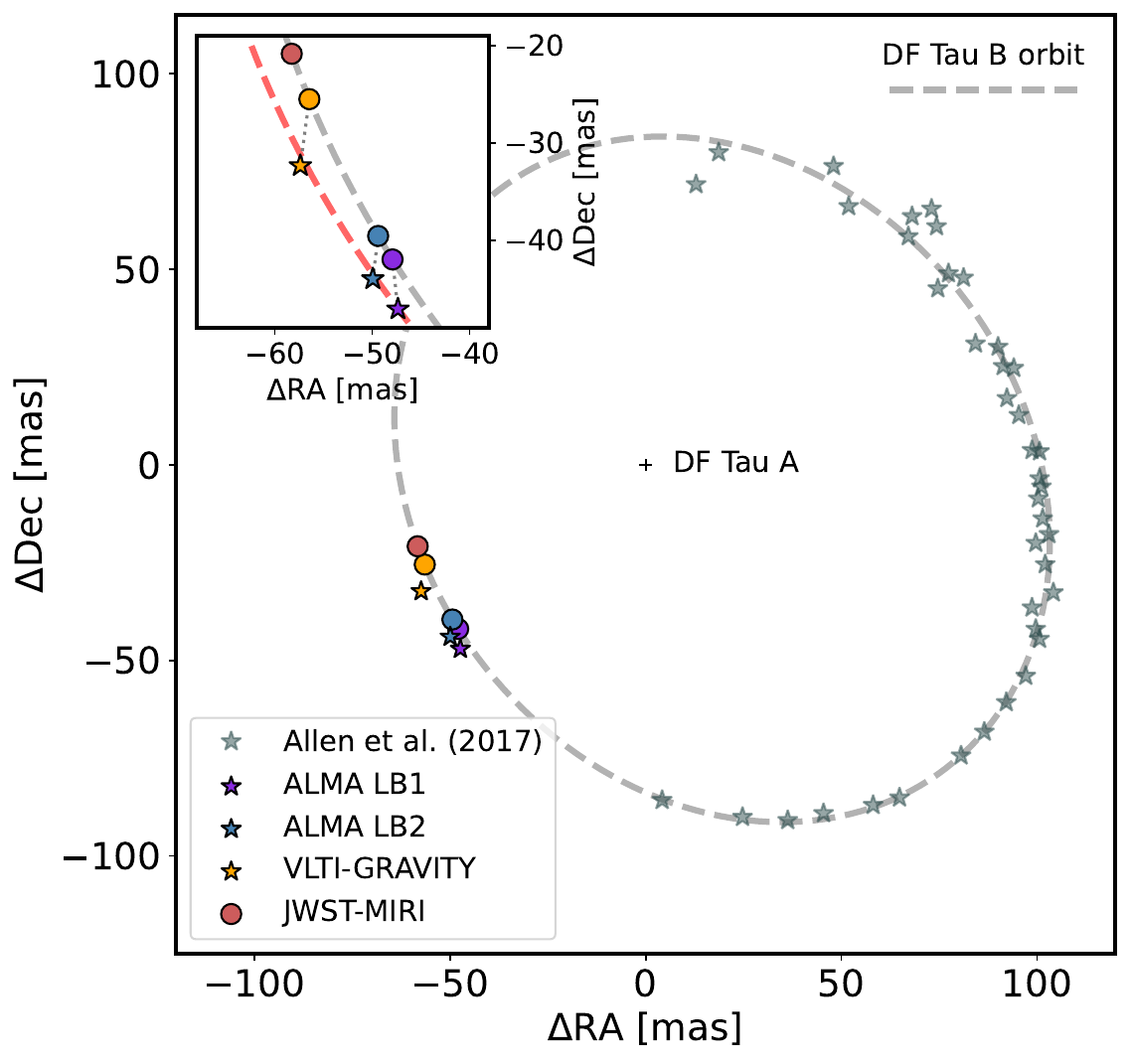}
    \caption{Astrometric measurements of DF Tau B relative to DF Tau A ($+$ symbol). The best orbit of DF Tau B from \citet{allen17} is shown with a dashed gray line, while the data points used to fit that orbit are shown with gray stars. The positions of DF Tau B relative to A measured with ALMA and VLTI-GRAVITY are shown with colored stars, while the solid circles show their expected position using the \citet{allen17} orbit. The offset from the \citet{allen17} orbit is within the orbital uncertainties, as shown by the red curve in the inset, which is a manual fit to the ALMA and VLTI-GRAVITY points that is within the uncertainties on the orbital parameters.}
    \label{fig:dftaub_astrometry}
\end{figure}

Both the VLTI-GRAVITY and ALMA interferometric measurements can be used to trace the relative position of DF Tau B to A (Figure~\ref{fig:dftaub_astrometry}). In the VLTI-GRAVITY observations, we find that the secondary, DF Tau B, is $57.4 \pm 0.2$ mas east of the primary and $32.4\pm 0.4$ mas to the south, with few differences between the two models used (see Section~\ref{subsec: gravity}). This translates to a separation $\rho=65.9\pm 0.6$ mas.

For the ALMA observations, we assumed that the star is located at the center of each emission blob, which we recovered using the visibility model in Section \ref{sec:alma_analysis}. We detect slight offsets between the expected location of DF Tau B based on the orbit from \citet{allen17} and our measured positions; however, these offsets are within the 1$\sigma$ variations of the orbital parameters (see the red line in the inset in Figure~\ref{fig:dftaub_astrometry}). An updated orbit determined from new observations will be presented in Kutra et al., in prep. 

The detection of movement between the ALMA LB1 and LB2 observations confirms that the secondary blob in the ALMA images is comoving with DF Tau B, and thus is likely a disk around DF Tau B. The projected distance between these measurements is about 3\,mas, or 0.5\,au at the distance of DF Tau B, and it is over 10 times smaller than the angular resolution of observations LB1 and LB2, which were taken only 60 days apart. This distance is as small as the pixel size of typical ALMA images of high angular resolution observations \citep[e.g.,][]{andrews18}, and it showcases the potential of using parametric visibility models to recover highly accurate astrometry of protoplanetary disks (see \citealt{kurtovic_sub} for a full discussion of using visibility modeling of ALMA datasets to recover astrometry of binary systems).

\section{Constraining the chemistry in the inner disk(s) with JWST-MIRI}\label{sec: JWST analysis}
\subsection{Slab modeling procedure}\label{subsec: slab models}

\begin{figure*}
    \centering
    \includegraphics[scale=0.5]{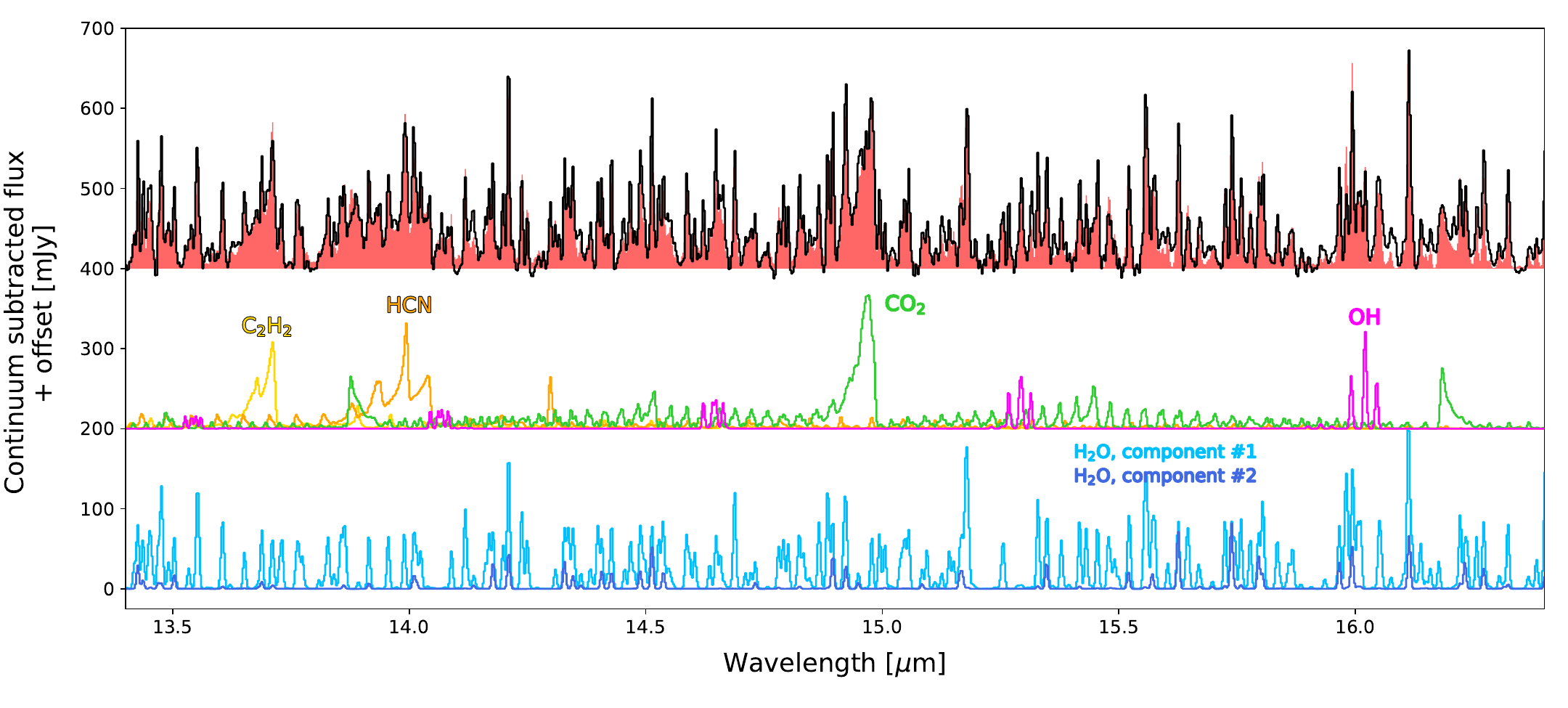}
    \caption{Continuum-subtracted MIRI spectrum (black) compared to the total model (red). Emission from C$_2$H$_2$ (yellow), HCN (orange), CO$_2$ (green), and OH (pink) is present. The two H$_2$O models are shown in light and medium blue. While this wavelength region is fit out to 18 \mic, we only show the spectrum out to 16.4 \mic\ for a better visualization of the main molecular species. }
    \label{fig: 13 to 18 mic}
\end{figure*}

Molecular emission from CO, H$_2$O, C$_2$H$_2$, HCN, CO$_2$, and OH is detected in the JWST-MIRI MRS spectrum of DF Tau (Figure~\ref{fig: spec}). Before we analyzed the spectrum, we subtracted the continuum. We followed the methods of \citealt{temmink24a}. Briefly, this involves an iterative process in which the continuum level is first determined using a Savitzky-Golay filter with a third-order polynomial. Emission lines over 2$\sigma$ above the continuum are masked so as to not skew the continuum estimation. The continuum is then subtracted, and all downward spikes more than 3$\sigma$ below the continuum are masked. Finally, the baseline is determined using \texttt{PyBaselines} \citep{erb22}.

Local thermodynamic equilibrium (LTE) slab models have been used to successfully reproduce the emission seen in JWST-MIRI observations (e.g., \citealt{grant23a,gasman23b,kospal23,perotti23,ramirez-tannus23,tabone23}), and therefore we used the same methods to analyze the DF Tau MRS data. In short, the slab models are computed with a Gaussian line profile with a FWHM of $\Delta$V=4.7 km s$^{-1}$ ($\sigma$=2 km s$^{-1}$) as is done in \cite{salyk11a} and mutual line shielding is accounted for for each molecular species. We did not take overlap between molecular species into account  in this analysis. The models only include three free parameters: the line-of-sight column density $N$, the gas temperature $T$, and the emitting area given by $\pi$$R^{2}$ for a disk of emission with radius $R$. We included emission from C$_{2}$H$_{2}$, HCN, H$_{2}$O, $^{12}$CO$_{2}$, and OH. The emitting area was varied to fit the spectrum, after being scaled to the appropriate distance. Given the binarity of this system, the \textit{Gaia} parallax is unreliable, in particular given that its renormalized unit weight error is nearly 22, which is much higher than the standard cutoff (1.4) for reliable parallaxes \citep{gaia21}. Therefore, we adopted the standard distance to Taurus of 140$\pm$10 pc, in agreement with the group location found in \cite{krolikowski21} and very similar to the weighted mean of \textit{Gaia} distances for stars within 30 arcminutes (135.66$\pm$3.22 pc, \citealt{akeson19}). Finally, the model spectrum is convolved to a resolving power that matches the observations and resampled to have the same wavelength sampling as the observed spectrum. We selected two spectral regions, from 13.4 to 18 \mic\ and from 18 to 25.6 \mic, where the resolving power is approximately constant across the two regions (subbands 3B to 3C and 4A to 4B, $R\sim$2500 and 1600, respectively; \citealt{labiano21, argyriou23}). The ro-vibrational water lines below $\sim$9 \mic\ are dominated by high energy ro-vibrational lines that can be suppressed due to non-LTE excitation (e.g., \citealt{banzatti23a,munoz-romero24a,pontoppidan24a}). As such, we did not fit the ro-vibrational lines at the same time as the rotational line at wavelengths longer than 10 \mic. We did not analyze the CO lines detected in the MIRI observations, as the wavelength coverage does not allow for a proper characterization of the CO properties (e.g., \citealt{grant24a, francis24, temmink24a}). While full CO analysis can be done with the IRTF-iSHELL data, this is outside the scope of this work. Beyond $\sim$26 \mic\ the noise level increases and artifacts are introduced due to the low signal at the long wavelengths in the flux reference star HD 163466 \citep{gasman23a}.\ We therefore limited our analysis to wavelengths shorter than this. 

While an iterative fitting method has done well reproducing JWST-MIRI spectra in some objects (i.e., fitting one molecule, subtracting it off then fitting the next molecule, etc.; \citealt{grant23a}), the strong forest of water lines across the DF Tau spectrum and the unique nature of this system led us to adopt a different methodology, as the spectrum is so line-rich that there is too much overlap between species to allow for a good fit using that method. Instead, we allowed for multiple slab components to be present in our model fitting and used an MCMC fitting procedure to fit all of the molecules and all components simultaneously, thereby reducing the effects of contamination by one species on the fit of another. C$_2$H$_2$ and HCN are only fit from 13.4 to 18 \mic, as they do not have strong contributions at longer wavelengths. Conversely, the third H$_2$O component, which is the coldest component, is only sensitive to the longest wavelengths; therefore, it is only constrained using the 18 to 25.6 \mic\ spectral range.
All other molecules and components are fit simultaneously across the full 13.4 to 25.6 \mic\ range. The fitting is done using \texttt{emcee}, using a uniform prior for each parameter, and eight times the number of walkers compared to the number of free parameters. We determine a noise level of 2 mJy in the 13.4 to 18 $\mu$m region (determined from the small line-free region from 16.34 to 16.36 $\mu$m) and a level of 4 mJy in the 18 to 25.6 $\mu$m region (determined from 20.59 to 20.62 $\mu$m), which is used in the fitting. This noise level corresponds to a signal-to-noise ratio between $\sim200$ and $450$ at these wavelengths. Model convergence was achieved after $\approx10^5$ steps. The posterior distributions are shown in Figure~\ref{fig: corner plot}. Finally, we note that OH is not in LTE (e.g., \citealt{tabone21}); therefore, we simply aimed to get a reasonably good fit such that the other species can be modeled without being impacted by the OH emission. The best-fit parameters for OH should then be taken with caution.

\begin{figure*}
    \centering
    \includegraphics[scale=0.5]{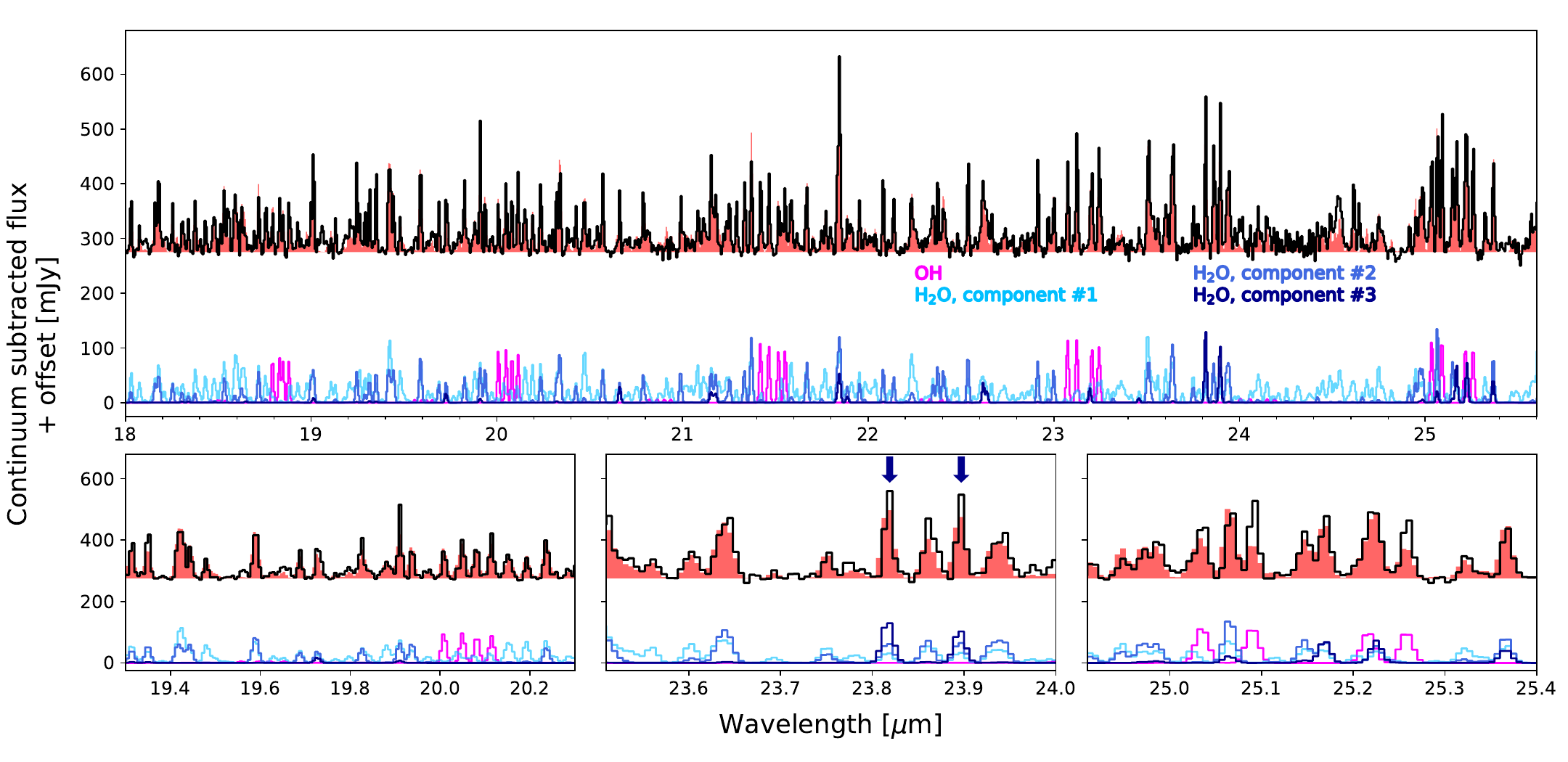}
    \caption{Continuum-subtracted MIRI spectrum (in black) compared to a total model (in red). The three water models and the OH fit are shown in colors below. The insets show the fits at different wavelengths. The two blue arrows indicate the two water lines discussed in \citet{banzatti23b} that signal a cold, $\sim$200 K water component.}
    \label{fig: water}
\end{figure*}

\subsection{Best-fit models to the JWST-MIRI data}
\begin{figure*}
    \centering
    \includegraphics[scale=0.56]{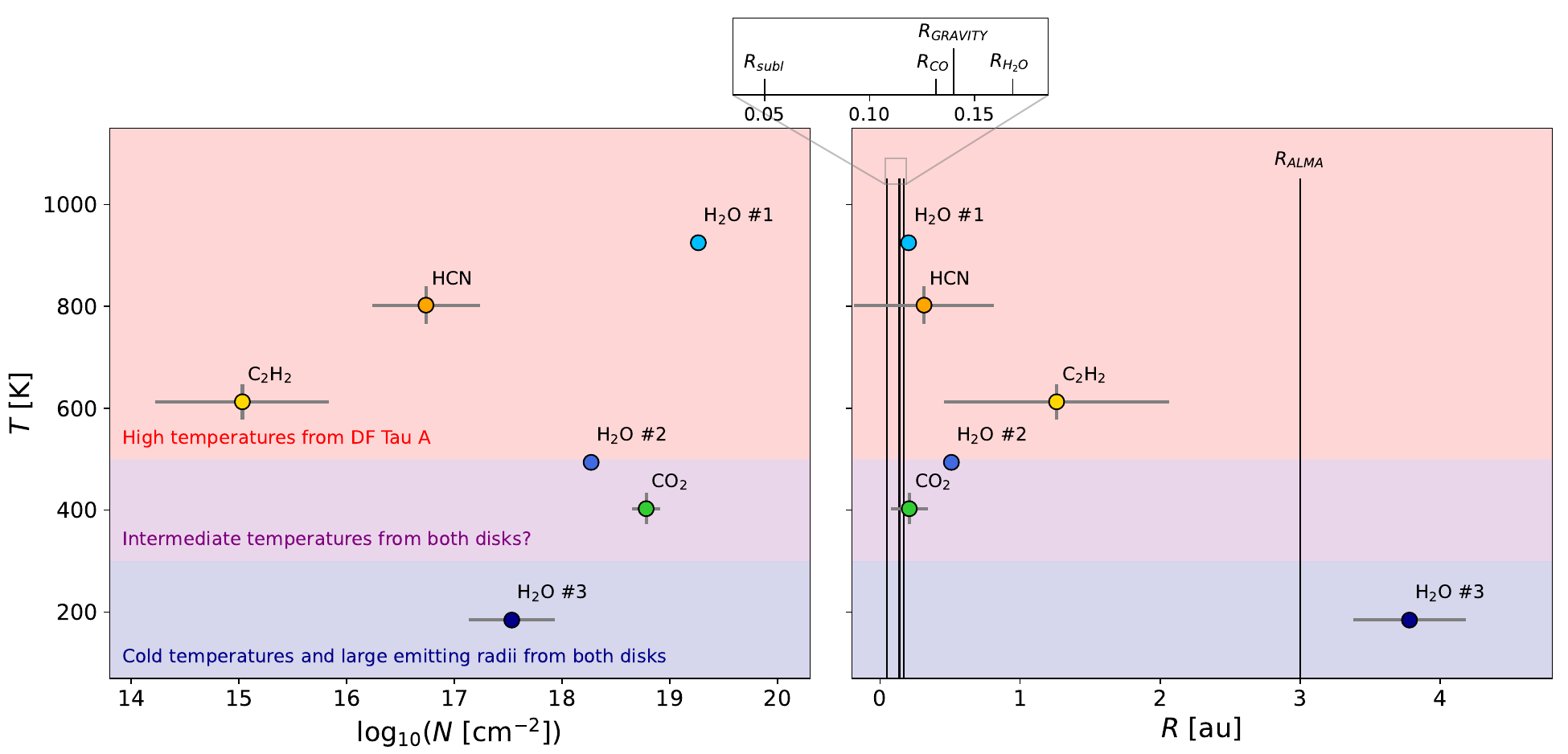}
    \caption{Best-fit parameters for the slab model fits to the DF Tau spectrum. Error bars, which are smaller than the points in some cases, are the 3$\sigma$ confidence intervals of the posterior distributions; the full posteriors are available in Figure~\ref{fig: corner plot}. The sublimation radii $R_{CO}$ (broad component) and $R_{H_2O}$ (derived from the IRTF-iSHELL observations) and the dust radii from VLTI-GRAVITY and ALMA are shown with the labeled black lines. As the JWST data are spectrally and spatially unresolved, only the emitting area can be determined; therefore, while the points for the molecular emission in the right panel may be true disk radii, they could also instead represent an annulus farther out with the same emitting area.}
    \label{fig: TN TR 13 to 25}
\end{figure*}

The best-fit model parameters are provided in Table~\ref{tab: model values} and these model spectra are compared to the observations in Figures~\ref{fig: 13 to 18 mic} and \ref{fig: water}, for the 13.4 to 18 $\mu$m region and the 18 to 25.6 $\mu$m region, respectively. The best-fit values for each species are shown in Figure~\ref{fig: TN TR 13 to 25}. Most of the emission is found to be coming from fairly high temperatures ($>$500 K), high column densities ($>$10$^{17}$ cm$^{-2}$), and small radii ($<$0.2 au). 

For most molecules, the emission is well reproduced by a single slab component; however, three components are needed to reproduce the water emission in the DF Tau spectrum, as seen in other sources as well (e.g., \citealt{gasman23b, banzatti23b}). We tested adding a fourth water component. The posteriors for this component clustered at an emitting radius of 0 au, however, indicating that three components are sufficient to reproduce the spectra in these wavelength ranges. Most species are in the optically thin regime; however, HCN appears to be at the border between optically thick and thin and the two hot water components and CO$_2$ are optically thick, which can be identified by the degeneracy between the column density and temperature (see Figure~\ref{fig: corner plot}). For the optically thin species or components, there is a degeneracy between the column density and the emitting radius. Most emission has a small emitting area, with an equivalent radius within 1 au. However, while the coldest water component region, being in the optically thin regime, is not well constrained, the emitting area is more extended than the other species. Two low-energy H$_2$O lines at 23.8 to 23.9 \mic, identified by \cite{banzatti23a} as being indicative of cold ($\lesssim$200 K) gas, are still under-predicted, indicating that even more cold water may be present. The water emission shows a decreasing temperature with increasing emitting area, as seen in other sources as well (e.g., \citealt{gasman23b}) and we discuss the origin of the cold ($\lesssim$ 200 K) water component in Section~\ref{subsec: cold water}. 

After analyzing the residuals, we do not find evidence for the detection of any isotopologues, such as $^{13}$CO$_2$ \citep{grant23a} or $^{13}$CCH$_2$ \citep{tabone23} that have been detected in other disks observed with JWST.

\begin{table*}
\centering
\caption{Best-fit JWST-MIRI model parameters.}
\label{tab: model values}
\begin{tabular}{ccccccc}
\hline \hline 
Species & log$_{10}(N)$ & $T$ & $R$ & $\mathcal{N}_{tot}$\\
 &   [cm$^{-2}$] &  [K] & [au] & [mol.] \\
\hline
H$_2$O \#1 & 19.26$^{+0.03}_{-0.03}$ & 920$^{+10}_{-10}$ & 0.20$^{+0.002}_{-0.002}$ & 5.3$\times 10^{44}$ \\ 
H$_2$O \#2 & 18.27$^{+0.02}_{-0.03}$ & 490$^{+10}_{-10}$ & 0.51$^{+0.011}_{-0.007}$ & 3.4$\times 10^{44}$ \\ 
H$_2$O \#3 & 17.53$^{+0.09}_{-0.31}$ & 180$^{+10}_{-0}$ & 3.78$^{+1.027}_{-0.262}$ & 3.4$\times 10^{45}$ \\ 
CO$_2$ & 18.78$^{+0.07}_{-0.06}$ & 400$^{+10}_{-20}$ & 0.21$^{+0.011}_{-0.009}$ & 1.8$\times 10^{44}$ \\ 
C$_2$H$_2$ & 15.03$^{+0.70}_{-0.11}$ & 610$^{+20}_{-20}$ & 1.26$^{+0.171}_{-0.689}$ & 1.2$\times 10^{42}$ \\ 
HCN & 16.73$^{+0.22}_{-0.29}$ & 800$^{+20}_{-20}$ & 0.31$^{+0.110}_{-0.061}$ & 3.8$\times 10^{42}$ \\ 
\hline
OH & 14.09$^{+0.49}_{-0.08}$ & 1880$^{+20}_{-20}$ & 3.16$^{+0.302}_{-1.344}$ & 8.7$\times 10^{41}$ \\ 
\hline
\end{tabular}
\tablefoot{While we provide the values for OH, we note that this molecule is not in LTE, and therefore the numbers do not accurately reflect the conditions of OH.}
\end{table*}

\section{Discussion}\label{sec: discussion}

\subsection{Does the disk around DF Tau B have a cavity?}\label{subsec: geometry of disk around B}

The discovery of the robust millimeter continuum from DF Tau B is surprising, given other spatially resolved observations at shorter wavelengths, which indicated no disk around this star (e.g., \citealt{schaefer14,allen17}). Given the marginally resolved nature of the ALMA data, it is unlikely that this is a flare event \citep{macgregor21}. One way to reconcile the lack of accretion signatures and infrared excess with the strong millimeter continuum is if there is a cavity in the inner disk. Such a cavity would remove hot ($\sim$1000-1500 K) dust and gas to reproduce the near-infrared spectral energy distribution (SED) while allowing an outer disk to emit at millimeter wavelengths (e.g., \citealt{espaillat14}). For DF Tau B, there is likely a ``sweet spot'' in terms of cavity size at which no near-infrared excess is present and that is not larger than what is allowed by the current ALMA observational limits. Constraining the cavity size is outside the scope of this work and may be impossible without higher angular resolution ALMA observations (for instance, even full SED modeling may be too degenerate given that no resolved photometry is available between 4 \mic\ and 1.33 millimeter). However, while we cannot say with certainty the size of this cavity, we suggest that a $\sim$1 au cavity may be possible in this disk. This is illustrated in Figure~\ref{fig: cartoon}.

\subsection{Hot emission from the disk around DF Tau A?}

Unlike DF Tau B, DF Tau A has a robust inner disk. This is inferred from its active accretion (\Mdot$\sim$1.7$\times$10$^{-8}$ \msunyr; scaled from the value in \citealt{gangi22} using the distance and stellar parameters from \citealt{allen17}) and the emission of gas and dust within 0.2 au, determined from IRTF-iSHELL and VLTI-GRAVITY, respectively. Given the stellar properties of DF Tau A, the sublimation radius is $\sim$0.05 au, which is determined from $R_{subl.} = \sqrt{L_*/4\pi\sigma T_{subl}^4}$ (following \citealt{banzatti22a}, similar to \citealt{monnier02,lazareff17,gravity21}), assuming a dust sublimation temperature of 1500 K and a stellar luminosity of 0.59 \Lsun\ (from \rstar=2.1 \Rsun\ and $T_{eff}$=3490 K; \citealt{allen17}). 

The radii from the spectrally and spatially resolved observations are shown in Figure~\ref{fig: TN TR 13 to 25} compared to the equivalent emitting radii from the model fits to the unresolved JWST data. Several complicating factors are at play in the interpretation of these results. First, the radii from the JWST data could either correspond to true emitting radii or instead represent an annulus at larger radii with the same emitting area. Second, without knowing the disk geometry of DF Tau B (i.e., the size of the cavity that may be present), it is unknown how much of the MIRI spectrum may be coming from DF Tau B. Detailed 2D thermochemical modeling in \cite{vlasblom24} showed that an inner disk cavity can produce bright mid-infrared spectral lines that emit at the cavity edge, and that the lines can be even brighter than a disk without a cavity, with the increase in flux depending on the cavity size and the stellar luminosity. With these limitations in mind, we can still make inferences about the origin of the mid-infrared gas emission, based on the temperature and, in some cases, the emitting area. For instance, from the models of \cite{vlasblom24}, we can estimate that the gas temperature at the emitting location of some of the mid-infrared spectral lines for a $\sim$0.6 \Lsun\ star and a 1 au cavity is $\sim$400 K (see their Figure B.3). While this is very approximate, it is useful as a benchmark, and indicates that much hotter gas can only originate from DF Tau A.

In this framework, we suggest that the two hottest H$_2$O components, HCN, and C$_2$H$_2$ emission are coming from the disk around DF Tau A, although the second H$_2$O component is borderline in terms of temperature. Interestingly, the radii determined from the fits to the MIRI data for the hot H$_2$O components and HCN match extremely well with the dust sublimation radius, the inner disk radius from VLTI-GRAVITY, and the CO and H$_2$O radii from IRTF-iSHELL, tentatively suggesting that these species could originate at this close-in location. While C$_2$H$_2$ is at larger radii, we note that C$_2$H$_2$ is completely optically thin; therefore, the degeneracy between the column density and emitting radius would allow the radius to be much smaller. 

While the emitting radius of CO$_2$ is very similar to the high temperature species and the radii of the spatially and spectrally resolved inner disk, the temperature is low enough that its origin is unclear. The CO$_2$ parameters are remarkably similar to those determined for GW Lup \citep{grant23a} with the hot-bands at 13.9 and 16.2 \mic\ being indicative of high column densities. The cool temperature ($\lesssim$400 K) has also been seen in other disks, like DR Tau \citep{temmink24a}, SY Cha \citep{schwarz24}, PDS 70 \citep{perotti23}, and Sz 98 \citep{gasman23b}, although higher temperatures ($\sim$700 K) were found to be most common in large \textit{Spitzer} samples \citep{salyk11a}. The low CO$_2$ temperature in the DF Tau system can indicate that the emission is coming from deeper in the disk or at larger radii than other molecules, or it could be a contribution from DF Tau B, potentially at the cavity edge.

\subsection{The origin of the cold water component}\label{subsec: cold water}

\begin{figure*}[t]
    \centering
    \includegraphics[scale=0.5]{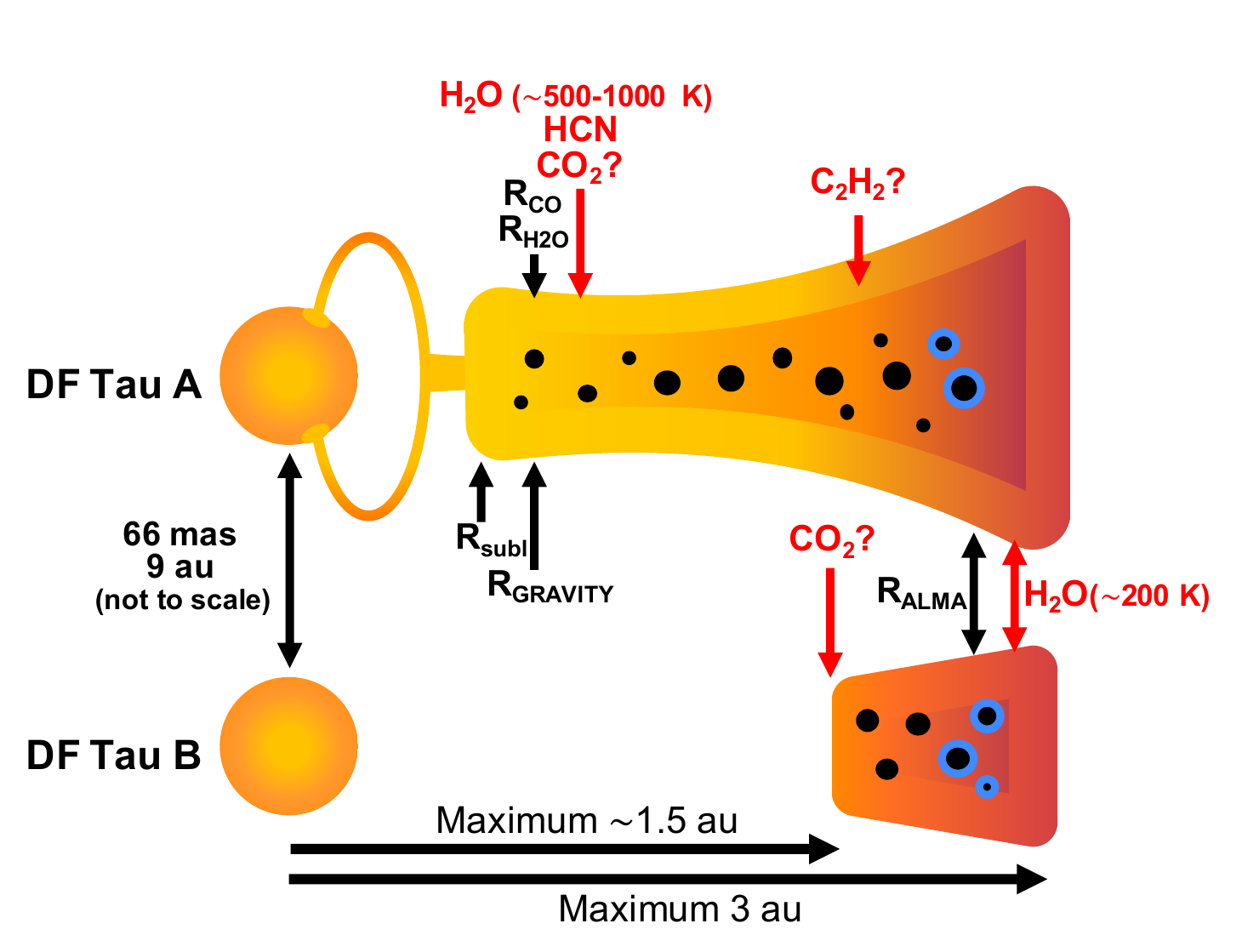}
    \caption{Illustration of the potential DF Tau A and B disk geometries. Radii determined from spectrally or spatially resolved emission are indicated in black. Emitting radii of the molecular emission observed with JWST are indicated in red. }
    \label{fig: cartoon}
\end{figure*}

There are three main formation pathways for water: gas-phase chemistry at temperatures above $\sim$250 K, ion-molecule chemistry at low temperatures, and on the surface of dust grains forming water-ice mantles (e.g., \citealt{vandishoeck21}). The temperature gradients in disks naturally make for differing water reservoirs, seen in both observations and modeling (e.g., \citealt{glassgold09, woitke09, blevins16, banzatti23b, gasman23b}). The hot water observed in the DF Tau MIRI observations is likely produced by gas-phase formation at high temperatures, possibly in the disk around DF Tau A, given the presence of hot, close-in disk material, and the presumed lack of such close-in material around DF Tau B as suggested by the absence of a near-infrared excess. The cold component, on the other hand, may be produced by water being sublimated directly from dust grains and going into the gas-phase, consistent with the extended emitting area needed to reproduce the spectrum. We discuss three potential pathways of this cold material, although we note that more than one origin is likely present. 

\subsubsection{Accretion activity}
The DF Tau system is dynamic and photometric and spectroscopic variability have long been observed in this system (e.g., \citealt{zaitseva76, johnskrull97, unruh98, allen17}), including what appeared to be an extreme flare from episodic mass accretion detected by a 6 magnitude variation in B-band photometry in a week-long period in 2000 \citep{li01}. Ground-based optical photometry of DF Tau, taken in the weeks surrounding our MIRI observations, showed increased variability around this time period (Kutra et al. in prep; priv. communication). 

An accretion outburst can push the snow line outward (e.g., \citealt{houge_krijt23, tobin23}). In DF Tau, an outburst could have moved the snow line outward, causing additional ice sublimation resulting in the cold water excess observed in the MIRI spectrum. Using the snow line calculation from \citet{mulders15} (their Equation 2) with the DF Tau A stellar properties of M$_{*}$=0.55 \Msun\ and an accretion rate of 1.7$\times$10$^{-8}$ \msunyr\ (see above), the snow line location in DF Tau A should be around 2.2 au. If the accretion rate increases by $\sim$0.5 dex, a change that is on the higher end but within the spread of accretion variability in other disks \citep{claes22}, the snow line will move out to over 3 au, larger than the expected size of the disk. Such a burst would thereby liberate all of the water ice from the grain surfaces. The very comparable locations of the outer dust disk and the snow line point to a low abundance of water ice and a high abundance of water vapor in this system.

\subsubsection{Radial drift}

Small-separation binarity can enhance radial drift (e.g., \citealt{zagaria21a,zagaria21b}). However, given the very small separation between DF Tau A and B, it is likely that the disks were never large; therefore, while enhanced radial drift is likely taking place in this system, there is not a very large reservoir of material for inward drift. Additionally, given that this migration would be the drift of material in this high density, very close-in region to the star (within 3 au), the particle sizes that can drift at these locations are on the order of boulders, not millimeter-sized grains. While this is possible, at these radii, radial drift would be extremely fast. In order to have enough material still emitting at millimeter wavelengths to be consistent with the ALMA observations, there would need to be a replenishment of $\sim$millimeter-sized grains, potentially via collisions, or a close-in dust trap keeping the dust disk from being fully accreted.

\subsubsection{An irradiated cavity wall in the disk of DF Tau B}\label{subsubsec: cavity in B}

Given that there are no accretion signatures for DF Tau B and there is no near-infrared excess in the SED, the disk around DF Tau B likely has a cavity. Such a cavity would remove the hot emission close to the star and provide a cavity edge to be directly irradiated. Recent 2D thermochemical modeling from \cite{vlasblom24} show that a cavity in a disk around a \Lstar=0.4 \Lsun\ star can produce strong mid-infrared water lines and can enhance the line fluxes in low energy lines, consistent with what we observe here in the DF Tau spectrum (see the inset in Figure~\ref{fig: water}). An illustration of the system with this configuration is shown in Figure~\ref{fig: cartoon}.

If a cavity is determined to be present in DF Tau B using higher angular resolution ALMA observations, we would be able to perform 2D thermochemical modeling to aid in the interpretation of the JWST spectrum (e.g., studying the impact of cavity size on the mid-infrared spectrum, as in \citealt{vlasblom24}).

\subsection{Comparison to other sources observed with JWST}
The emission of CO$_2$, HCN, and the second component of H$_2$O have similar properties to those found for GW Lup \citep{grant23a}, a T Tauri star with similar stellar properties (M1.5 spectral type, \mstar=0.46 \Msun, \Lstar=0.33 \Lsun) but very different outer disk properties given that the outer dust disk of GW Lup has a radius of $\sim$100 au \citep{dullemond18}, compared to the 3\,au of DF Tau A and B. However, DF Tau clearly has more complex H$_2$O emission, with the presence of the very hot and very cold components, which are not present in GW Lup. In this way, DF Tau is more similar to other water-rich disks (Sz 98, \citealt{gasman23b}; DR Tau, \citealt{temmink_sub}; GK Tau and HP Tau, \citealt{banzatti23b}). Given the unique nature of this binary system, the fact that multiple components are needed to reproduce the combined DF Tau A and B spectrum is not surprising; in fact, given this complexity, it is remarkable that only one component is needed for C$_2$H$_2$, HCN, and CO$_2$. Therefore, despite the very different physical structure of DF Tau A and B relative to previously studied sources, the inner disk chemistry remains relatively unchanged, indicating that the broader physical structure and outer disk evolution or properties are not driving the conditions in the innermost disk(s) in this system.

\section{Summary and conclusions}\label{sec: summary}
We present new JWST-MIRI MRS observations of the equal-mass, close-separation ($\sim$66 milliarcseconds, 9 au at the time of the observations presented here) DF Tau binary system. Pairing these new data with complementary ground-based data from VLTI-GRAVITY, ALMA, and IRTF-iSHELL, we have gained new insight into this unique system. 

\begin{enumerate}
    \item We find a previously unexpected cold dust disk around DF Tau B. High-resolution ALMA observations show a robust continuum detection at the location of DF Tau B. Precise astrometry with both ALMA and VLTI-GRAVITY shows movement of DF Tau B relative to DF Tau A, consistent with the previously published orbital motion. Movement is detected on timescales as short as 60 days. Previous high angular resolution photometry and spectroscopy in the UV, optical, and near-infrared indicated no disk around this star, unlike around DF Tau A. To reconcile the lack of near-infrared excess with the presence of millimeter continuum emission, we suggest that there may be a small ($\sim$1 au) cavity in this disk. With higher angular resolution ALMA observations, we would be able to determine if this cavity is indeed present and, if so, determine the cavity size.

    \item The JWST-MIRI MRS spectrum of the DF Tau system is extremely rich in molecular emission. Features of CO, C$_2$H$_2$, HCN, CO$_2$, and OH and a forest of strong H$_2$O lines are all clearly detected. DF Tau A and B are not spatially or spectrally resolved in the MIRI observations, and therefore the spectrum likely has contributions from both disks. 

    \item Multiple temperature components are needed to reproduce the rotational water lines observed in the spectrum, ranging from temperatures above 900 K down to a temperature below 200 K, with the latter coming from a very extended emitting area. This cold, extended emission is coming from an area larger than what is expected for either disk in this system and so must therefore originate in both disks. The hot water may be coming from the inner disk around DF Tau A, and the cold water may be coming from water ice sublimating off of the surfaces of dust grains. This cold water enrichment may be produced by accretion outbursts, enhanced radial drift due to the binarity, and/or an irradiated cavity wall around DF Tau B. 

    \item Overall, the molecular emission is relatively similar to that of disks around similar-type stars in isolated systems, despite the fact that the disk properties -- in particular the disk sizes -- are so different. This may indicate that the inner disk chemistry is largely independent of the outer disk evolution and properties in this system. The water emission, however, is quite complex, as might be expected given a contribution from both disks.

\end{enumerate}

JWST observations are unveiling the composition and conditions in the inner regions of protoplanetary disks. As these investigations continue, observing and analyzing additional multiple-star systems, particularly at a range of separations and with systems resolved in observations, will be important for determining the impact of multiplicity on the chemistry and evolution in disks. 

Finally, we wish to emphasize the power of combining observations from multiple facilities in developing a global view of protoplanetary disks, especially in dynamic, complex systems like DF Tau.

\begin{acknowledgements}
We thank Lisa Prato, Taylor Kutra, Benjamin Tofflemire, and Catherine Espaillat for useful discussions. 

This work is based on observations made with the NASA/ESA/CSA James Webb Space Telescope. The data were obtained from the Mikulski Archive for Space Telescopes at the Space Telescope Science Institute, which is operated by the Association of Universities for Research in Astronomy, Inc., under NASA contract NAS 5-03127 for JWST. These observations are associated with program \#1282. The following National and International Funding Agencies funded and supported the MIRI development: NASA; ESA; Belgian Science Policy Office (BELSPO); Centre Nationale d’Etudes Spatiales (CNES); Danish National Space Centre; Deutsches Zentrum fur Luft- und Raumfahrt (DLR); Enterprise Ireland; Ministerio De Econom\'ia y Competividad; Netherlands Research School for Astronomy (NOVA); Netherlands Organisation for Scientific Research (NWO); Science and Technology Facilities Council; Swiss Space Office; Swedish National Space Agency; and UK Space Agency.

GRAVITY is developed in a collaboration by the Max Planck Institute for Extraterrestrial Physics, LESIA of the Paris Observatory and IPAG of Université Grenoble Alpes / CNRS, the Max Planck Institute for Astronomy, the University of Cologne, the Centro Multidisciplinar de Astrofisica Lisbon and Porto, and the European Southern Observatory. We would like to thank all the individuals who have contributed to build the GRAVITY instrument.

N.K. thanks the Deutsche Forschungsgemeinschaft (DFG) - grant 138 325594231, FOR 2634/2. E.v.D. acknowledges support from the ERC grant 101019751 MOLDISK and the Danish National Research Foundation through the Center of Excellence ``InterCat'' (DNRF150). T.H. and K.S. acknowledge support from the ERC Advanced Grant Origins 83 24 28. I.K., A.M.A., and E.v.D. acknowledge support from grant TOP-1614.001.751 from the Dutch Research Council (NWO). I.K. and J.K. acknowledge funding from H2020-MSCA-ITN-2019, grant no. 860470 (CHAMELEON). H.N. acknowledges support from the French National Research Agency in the framework of the ``investissements d'avenir'' program (ANR-15-IDEX-02). K.P. acknowledges the support from the French National Research Agency for the
project ``ANR-23-EDIR-0001-01''. M.T. and M.V. acknowledge support from the ERC grant 101019751 MOLDISK. V.C. acknowledges funding from the Belgian F.R.S.-FNRS. B.T. is a Laureate of the Paris Region fellowship program (which is supported by the Ile-de-France Region) and has received funding under the Marie Sklodowska-Curie grant agreement No. 945298. D.G. would like to thank the Research Foundation Flanders for co-financing the present research (grant number V435622N) and the European Space Agency (ESA) and the Belgian Federal Science Policy Office (BELSPO) for their support in the framework of the PRODEX Programme. D.B. and M.M.C. have been funded by Spanish MCIN/AEI/10.13039/501100011033 grants PID2019-107061GB-C61 and No. MDM-2017-0737. A.C.G. has been supported by PRIN-INAF MAIN-STREAM 2017 and from PRIN-INAF 2019 (STRADE). G.P. gratefully acknowledges support from the Max Planck Society. Support for F.L. was provided by NASA through the NASA Hubble Fellowship grant \#HST-HF2-51512.001-A awarded by the Space Telescope Science Institute, which is operated by the Association of Universities for Research in Astronomy, Incorporated, under NASA contract NAS5-26555.

\end{acknowledgements}

\bibliographystyle{aa}

\bibliography{biblio}

\clearpage

\begin{appendix}

\section{ALMA log and  visibility model}\label{sec:alma_analysis}

The observing log of the ALMA observations is provided in Table~\ref{tab:alma_obs_log}.

To maximize the recovery of information, the dust continuum emission was analyzed using parametric visibility modeling with the packages \texttt{galario} \citep{tazzari2018} and \texttt{emcee} \citep{emcee}. The parametric brightness distribution describes each disk of DF Tau as a Gaussian ring, following\begin{equation}
    I_k(r) \,=\, f_{k} \cdot \exp \left( -0.5 \left(\frac{r - r_k}{\sigma_k}\right)^2 \right)\,\text{,}
\end{equation}

\noindent where $k$ can describe the disk around DF Tau A or DF Tau B depending on the disk, $r_k$ is the radial distance from the disk center, $f_k$ is the amplitude of the emission, $r_k$ the center of the Gaussian, and $\sigma_k$ is the standard deviation. The disks can become centrally peaked if $r_k=0$, which is a solution allowed in our models. We allowed the disks to have different positions for each observation, and thus the relative distance between the emission of the disks is a result of our fitting process. Additionally, we included a flux scaling parameter, $\gamma_{obs}$, which accounts for small differences in the total flux of each observation due to different flux calibrators and slightly different wavelengths (see the frequency coverage in Table~\ref{tab:alma_obs_log}). As a flux reference we used SB2, as it is the observation with the shortest baselines. Therefore, $\gamma_{SB2}=1$, and the other observations are scaled to match it.

In Table \ref{appendix:tab:uv_mcmc} we show the additional parameters fitted in our visibility modeling to the ALMA data. The phase center given in the table refers to the distance of DF Tau A center to the center of the observation. The intensity scaling of SB2 is exactly 1.0, as this observation was used as a flux reference, and thus its flux amplitude is 1.0 of itself. The Gaussian ring used to fit each disk is a function of the radii $r$ starting from the center of each disk, following
\begin{equation}
    G(r) = f_{\text{disk}} \cdot \exp{\left( -0.5 \cdot \frac{(r-r_{\text{disk}})^2}{\sigma_{\text{disk}}^2} \right)} \text{,}
    \label{app:eq:Gaussian}
\end{equation}

\noindent where ``disk'' can be either A or B. This functional form allows the disks to be centrally peaked or ringed, based on their specific morphology.

\begin{table*}
\caption{ALMA observations of DF Tau included in this work. }
\centering
\begin{tabular}{ c|c|c|c|c|c|c|c } 
  \hline
  \hline
\noalign{\smallskip}
Project & Code & PI Name & Obs Date & N        & Baselines & Exp Time & Freq \\
code    &      &         &          & antennas & (m)       & (min)    & (GHz) \\
\noalign{\smallskip}
  \hline
\noalign{\smallskip}
2019.1.01739.S  & SB1 & Tofflemire, B. & 2021-07-18 & 49 & 15 - 3697   & 5.14  & 230.07 - 247.90 \\
                & LB1 &                     & 2021-08-24 & 48 & 47 - 12595  & 45.29 & \\
\noalign{\smallskip}
  \hline
\noalign{\smallskip}
2021.1.00854.S  & LB2 & Long, F.          & 2021-10-29 & 45 & 64 - 8283   & 26.31 & 217.60 - 233.93\\
                & SB2 &                     & 2022-07-04 & 42 & 15 - 1997   & 6.25  & \\
\noalign{\smallskip}
  \hline
  \hline
\end{tabular}
\label{tab:alma_obs_log}  
\end{table*}

\begin{table}[t]
\caption{Visibility modeling results, in addition to the parameters given in Table \ref{tab:uv_mcmc}.}
\centering
\begin{tabular}{ c|c|c|c } 
  \hline
  \hline
\noalign{\smallskip}
           & Property      & Best value $\pm 3\sigma$ & unit \\
\noalign{\smallskip}
  \hline
\noalign{\smallskip}
Phasecenter  & $\Delta\text{RA}_{\text{SB1}}$  & $-24.2 \pm 1.7 $  & mas \\
DF Tau A     & $\Delta\text{Dec}_{\text{SB1}}$ & $-24.6 \pm 1.7 $  & mas \\
             & $\Delta\text{RA}_{\text{SB2}}$  & $ 25.1 \pm 9.7 $  & mas \\
             & $\Delta\text{Dec}_{\text{SB2}}$ & $-12.2 \pm 14.3 $ & mas \\
             & $\Delta\text{RA}_{\text{LB1}}$  & $ 27.5 \pm 0.1 $  & mas \\
             & $\Delta\text{Dec}_{\text{LB1}}$ & $-29.5 \pm 0.2 $  & mas \\
             & $\Delta\text{RA}_{\text{LB2}}$  & $ 24.8 \pm 0.6 $  & mas \\
             & $\Delta\text{Dec}_{\text{LB2}}$ & $-21.3 \pm 1.1 $  & mas \\
\noalign{\smallskip}
  \hline
\noalign{\smallskip}
Intensity    & $i_{\text{SB1}}$ & $1.20 \pm 0.04$ & - \\
Scaling      & $i_{\text{SB2}}$ & $1.00$ & - \\
             & $i_{\text{LB1}}$ & $1.10 \pm 0.04$ & - \\
             & $i_{\text{LB2}}$ & $0.92 \pm 0.03$ & - \\
\noalign{\smallskip}
  \hline
\noalign{\smallskip}
Gaussian     & $f_{\text{A}}$      & $55.3_{-8.0}^{+3.3}$  & $\mu$Jy/pix \\
ring         & $r_{\text{A}}$      & $0.1_{-0.1}^{+3.2}$   & mas \\
             & $\sigma_{\text{A}}$ & $9.8_{-1.3}^{+0.3}$   & mas \\
             & $f_{\text{B}}$      & $44.5_{-3.3}^{+12.5}$ & $\mu$Jy/pix \\
             & $r_{\text{B}}$      & $4.1_{-4.1}^{+2.2}$   & mas \\
             & $\sigma_{\text{B}}$ & $8.3_{-1.1}^{+1.7}$   & mas \\
\noalign{\smallskip}
  \hline
\noalign{\smallskip}
\emph{Disk}           & inc$_A$ & $18.9_{-9.4}^{+6.0}  $ & deg \\
\emph{Geometry}$^{*}$ & PA$_A$  & $34.4_{-20.0}^{+31.9}  $ & deg \\
                      & inc$_B$ & $26.5_{-8.1}^{+4.7}  $ & deg \\
                      & PA$_B$  & $ 4.0_{-4.0}^{+15.0}  $ & deg \\
\noalign{\smallskip}
  \hline
  \hline
\end{tabular}
\tablefoot{$^{*}$ indicates parameters that should be taken with caution. As the emission is only marginally resolved, the values are not well constrained and the uncertainties instead reflect the uncertainties of the model and not the true uncertainties of the data.  }
\label{appendix:tab:uv_mcmc}
\end{table}

\section{The spectral energy distribution}\label{sec: sed}

\begin{figure*}
    \centering
    \includegraphics[scale=0.64]{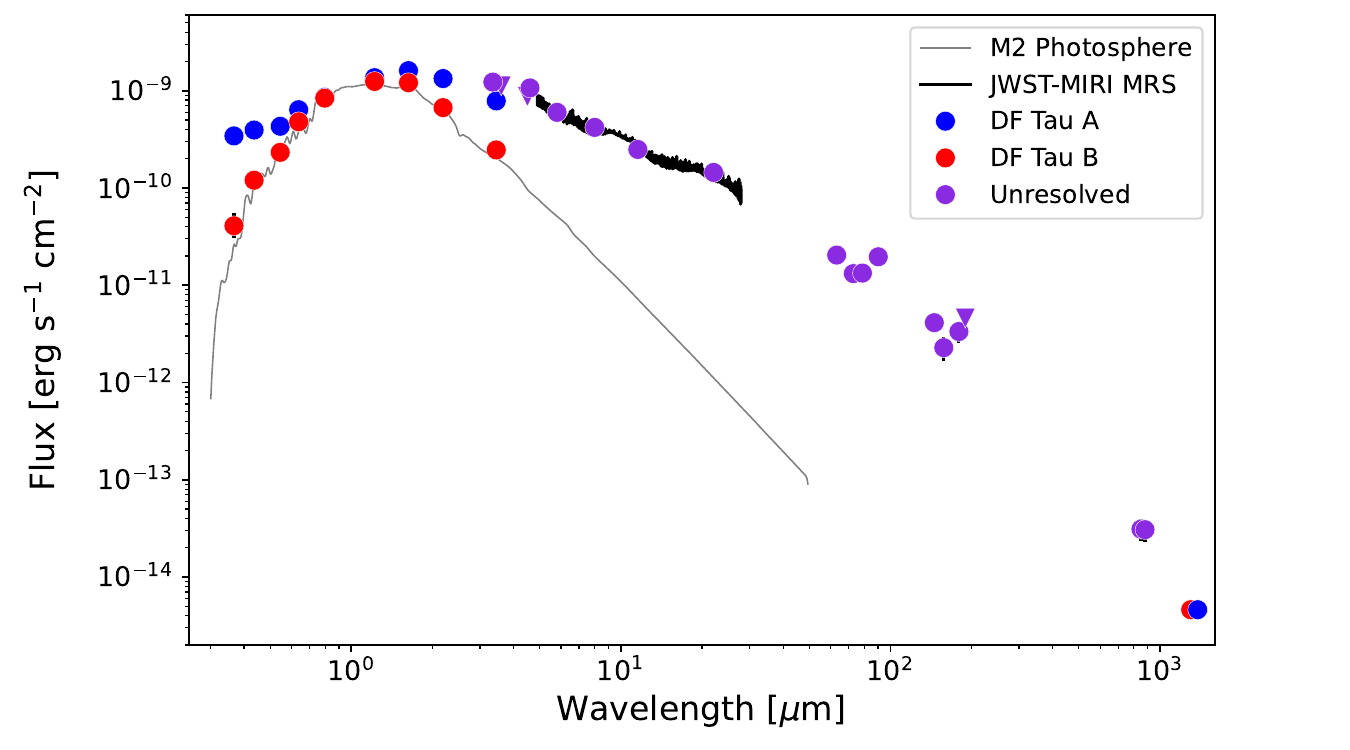}
    \caption{ SED for DF Tau A (blue) and B (red). Photometry that does not resolve A and B is shown in purple. The JWST-MIRI MRS spectrum is shown in black. A BT Settl model photosphere for an M2-type star is shown in gray, scaled to the J-band photometry, for reference \citep{allard12}. The SED has been de-reddened with an $A_V$=0.5 mag \citep{allen17}. Error bars are smaller than the markers in most cases. The photometric point at 1.3 mm from ALMA is the same for DF Tau A and B and the point for DF Tau A is shifted horizontally for clarity.}
    \label{fig: sed}
\end{figure*}

The SED for DF Tau is shown in Figure~\ref{fig: sed}. Literature photometry was collected from \cite[and references therein]{allen17} and \cite{howard13}. The 1.3 mm points are the ALMA data presented in this paper. A BT Settl photospheric model is shown for an M2-type star. The UV excess and near-infrared excess is clearly absent for DF Tau B, which is a signature of inner disk clearing (e.g., \citealt{espaillat14}). 

\section{VLTI-GRAVITY model}\label{sec: gravity model}
\begin{figure*}
    \centering
    \includegraphics[scale=0.54]{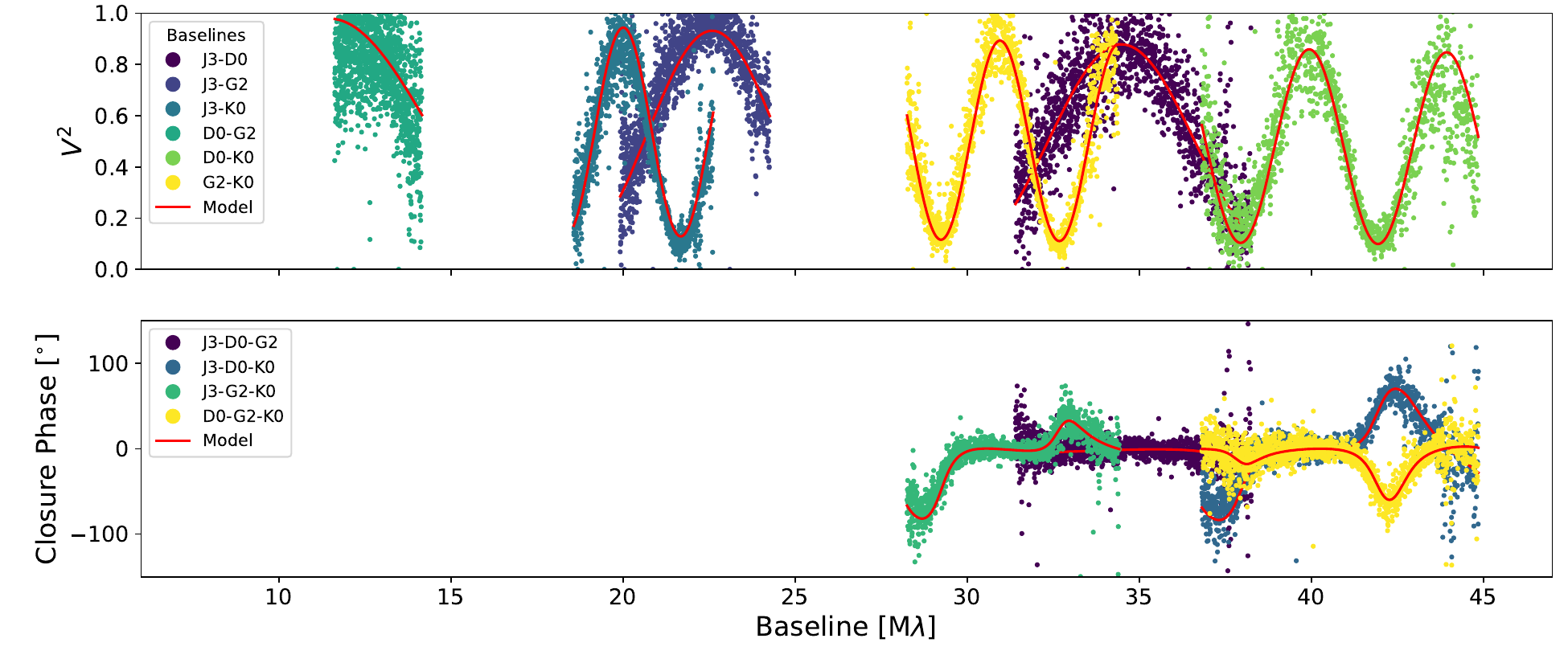}
    \caption{ VLTI-GRAVITY visibilities squared (top) and the closure phases (bottom) in the observations (colors) compared to the best-fit model (red).}
    \label{fig: gravity model}
\end{figure*}

Due to the marginally resolved nature of the VLTI-GRAVITY data, in order to determine the size of the inner disk, the flux must be constrained. To do this, we computed a synthetic K-band flux given a stellar photosphere with a temperature of 3500 K, in line with the spectral type of DF Tau A, and subtracted that flux from the observed value, resulting in a disk-only flux. This gives a ratio of the flux of the primary ($F_A$) to that of the disk ($F_{disk}$) of $F_A/F_{disk}= 6.09$. We then added a normalization such that $F_A+F_B+F_{disk}=1$, where $F_B$ is the flux of the secondary. This leaves us with a single free parameter to adjust the fluxes, and the flux-size degeneracy is broken. The best-fit model for the VLTI-GRAVITY data is shown compared to the observations in Figure~\ref{fig: gravity model}. The sinusoidal variations in the visibilities squared is a signature of binarity.

\section{The JWST-MIRI model results}\label{sec: corner plots}
\begin{figure*}
    \centering
    \includegraphics[scale=1.]{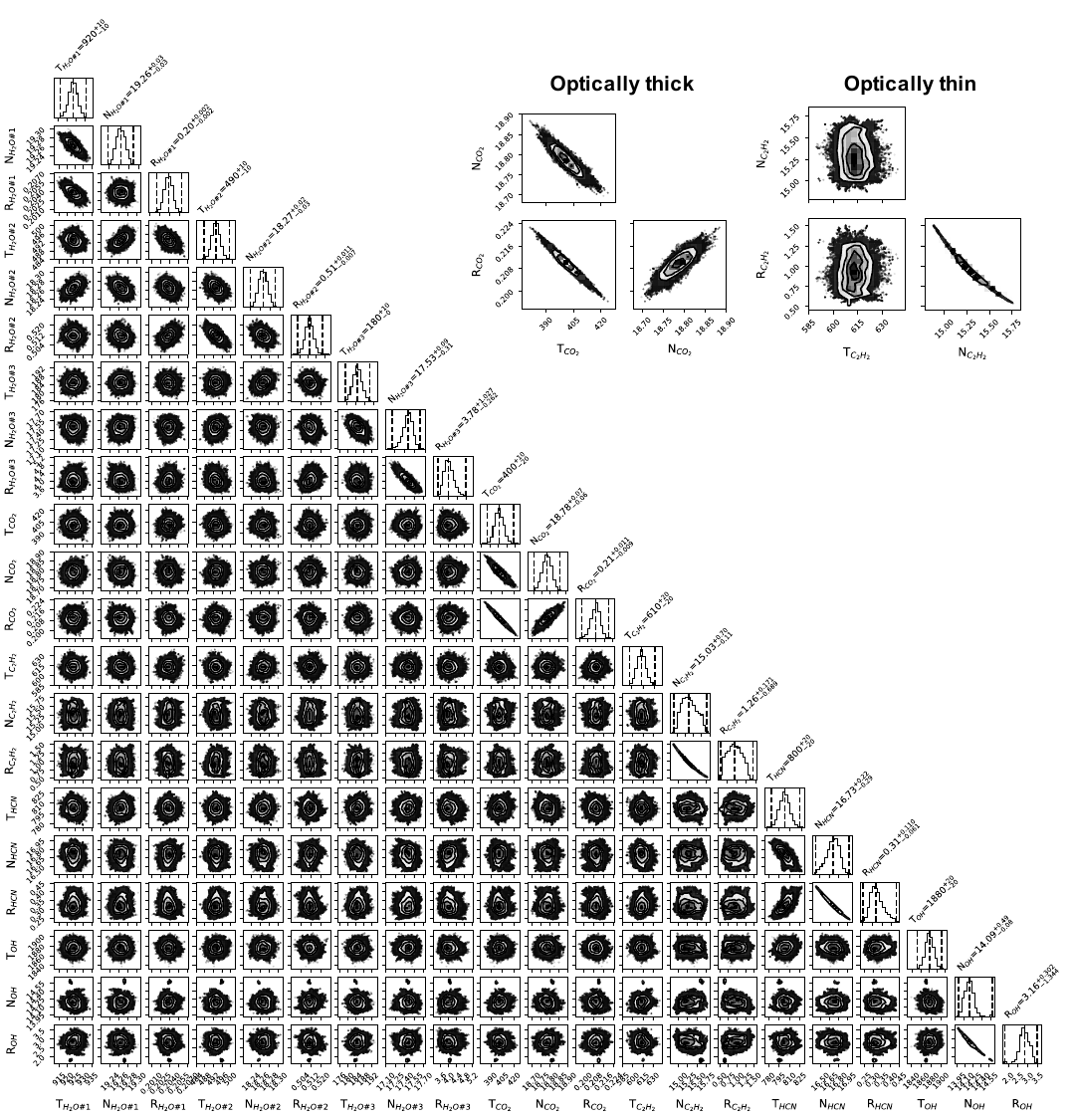}
    \caption{Posterior distributions of the MCMC modeling of the JWST-MIRI spectra. The column densities, $N$, are in log$_{10}$ space. Examples of the correlations seen in optically thick and optically thin cases are shown in the upper right. }
    \label{fig: corner plot}
\end{figure*}

The posterior distributions from the MCMC fitting to the JWST-MIRI observations are shown in Figure~\ref{fig: corner plot}. The best-fit values and the 3$\sigma$ confidence uncertainties are given above the histograms.

\end{appendix}

\end{document}